\documentclass[12pt,preprint]{aastex}

\newcommand{\msol}{\mathrm{M}_\odot}
\newcommand{\degree}{\ensuremath{^\circ}}

\slugcomment{To appear in ApJ}
\usepackage{longtable}

\shorttitle{A Rich Young Cluster, Cep OB3b}
\shortauthors{Allen et al.}

\begin{document}

\title{{\it Spitzer} Imaging of the Nearby Rich Young Cluster, Cep OB3\lowercase{b}}

\author{Thomas S. Allen\altaffilmark{1}, Robert
  A. Gutermuth\altaffilmark{2}, Erin Kryukova\altaffilmark{1},
  S. Thomas Megeath\altaffilmark{1}, Judith L. Pipher\altaffilmark{3},
  Tim Naylor\altaffilmark{4}, R. D. Jeffries\altaffilmark{5}, Scott J. Wolk\altaffilmark{6}, Brad Spitzbart\altaffilmark{6}, James
  Muzerolle\altaffilmark{7} }

\altaffiltext{1}{University of Toledo, Department of Physics and Astronomy, Toledo OH 43606}
\altaffiltext{2}{Department of Astronomy, University of Massachusetts, Amherst, MA 01003}
\altaffiltext{3}{Department of Physics and Astronomy, University of Rochester, Rochester, NY 14627}
\altaffiltext{4}{School of Physics, University of Exeter, Exeter, UK EX4 4QL}
\altaffiltext{5}{Astrophysics Group, School of Physical and Geographical Sciences, Keele University, Keele, Staffordshire, UK ST5 5BG}
\altaffiltext{6}{Harvard-Smithsonian Center for Astrophysics, Mail Stop 42, 60 Garden Street, Cambridge, MA 02138}
\altaffiltext{7}{Space Telescope Science Institute, Baltimore, MD 21218}

\begin{abstract}
We map the full extent of a rich massive young cluster in the Cep OB3b
association with the IRAC and MIPS instruments aboard the {\it
  Spitzer} Space Telescope\footnote{This work is based [in part] on
  observations made with the Spitzer Space Telescope, which is
  operated by the Jet Propulsion Laboratory, California Institute of
  Technology under a contract with NASA.} and the ACIS instrument aboard the $\it{Chandra}$ X-Ray Observatory\footnote{This work is based [in part] on observations made by the Chandra X-ray Observatory Center, which is operated by the Smithsonian Astrophysical Observatory for and on behalf of the NASA}.  At 700 pc, it is revealed
to be the second nearest large ( $>1000$ member), young ($< 5$~Myr)
cluster known.  In contrast to the nearest large cluster, the Orion
Nebula Cluster, Cep OB3b is only lightly obscured and is mostly
located in a large cavity carved out of the surrounding molecular
cloud.  Our infrared and X-ray datasets, as well as visible photometry from the literature, are used to take a census of the young stars in Cep OB3b.  We find that the young stars within the cluster are concentrated in two sub-clusters; an eastern sub-cluster, near the Cep B molecular clump, and a western sub-cluster, near the Cep F molecular clump.  Using our census of young stars, we examine the fraction of young stars with infrared
excesses indicative of circumstellar disks.  We create a map of the disk fraction throughout the cluster and find that it is spatially variable.
Due to these spatial variations, the two sub-clusters exhibit substantially different average disk fractions from each other: $32\% \pm 4\%$ and $50\% \pm 6\%$.  We discuss whether the discrepant disk fractions are due to the photodestruction of disks by the high mass members of the cluster or whether they result from differences in the ages of the sub-clusters. We conclude that the discrepant disk fractions are most likely due to differences in the ages.

\end{abstract}


\section{INTRODUCTION}

Surveys of molecular clouds show that approximately half of low mass
stars form in rich, massive clusters with more than a thousand young
stars \citep{carpenter00, dewitt05, allen07}.  Most nearby young
clusters are of a small to moderate size, containing at most a few
hundred members \citep{gutermuth09}.  Consequently, much of what we
know about star formation in rich, massive clusters comes from
studying the nearest example: the Orion Nebula Cluster (ONC) at a
distance of 414 pc \citep{menten07}. Here we use the term ONC to
denote not just the dense, $\sim$ 600 member cluster core found toward
the brightest regions of the Orion Nebula, but rather the extended
8~pc $\times$ 4~pc region containing $\sim 2000$ forming stars (Megeath et al. (in preparation); \citet{carpenter00}).

Cepheus OB3b was recognized as a distinct subgroup in the Cepheus OB3
association by \citet{blauuw64}.  Subsquent observations of the border
between the Cepheus OB3 molecular cloud and the Cep~OB3b subgroup
detected a young group of $>$ 400 low mass, pre-main sequence (PMS)
stars \citep{moreno93, jordi96, mikami01, pozzo03, getman06}.  Some of
these stars are embedded in the molecular clump Cep B
\citep{sargent77} near the HII region Sh 155 \citep{sharpless59};
however, most are lightly obscured and can be studied at visible
wavelengths.  \citet{jordi96} established an age of 5.5 - 7.5 Myr by
fitting theoretical isochrones to dereddened {\it V} and {\it B}
measurements for twenty-one stars in Cep~OB3b, considerably older than
the average age of the ONC \citep[$<2$ Myr,][]{hill97,jeff07}.  Based
on an analyis of the {\it V} vs. {\it V-I} color magnitude diagram,
\citet{mayne07} proposed that Cep~OB3b has an age of $\sim$3 Myr at an
adopted distance of 850 pc.  Using {\it Chandra} observations to
identify PMS members of Cep~OB3b, \citet{getman09} determined
extinctions to those stars with 2MASS near-IR color magnitude diagrams
and fit evolutionary isochrones to a de-reddened {\it V} vs. {\it V-I}
color-magnitude diagram of those stars.  They deduced an age of 2-3
Myr for the part of the cluster near Cep B using an adopted distance
of 725 pc.  More recent work by \cite{littlefair09} suggests an older
age estimate ($\sim$4.5 Myr, compared to an age of $2$~Myr for the ONC) with a distance of $580 \pm
60$~pc. Although the distance estimates in the literature vary from
580-850 pc, we adopt a distance of 700 pc; this is the maser
parallax-determined distance for Cepheus A which lies in the same
molecular cloud \citep{moscadelli08}.  In all of these studies, Cep~OB3b
appears to be older than the ONC. 

As part of a survey of the Cepheus OB3 molecular cloud we have mapped
the Cep~OB3b cluster with the IRAC and MIPS instruments aboard {\it
  Spitzer}.  When the {\it Spitzer} observations revealed the full extent of the cluster, we then mapped the cluster with the ACIS instrument aboard {\it Chandra}.  This is the first mid-infrared and X-ray census of young, low mass
stars in the {\it entire} Cep~OB3b cluster. Using these data, we map
the distribution of young stars with enhanced infrared (due to dusty circumstellar material) or X-ray (due to increased chromospheric activity) emission. These data show that Cep OB3b is one
of the largest known clusters within 1 kpc of the Sun.  We propose that
the cluster may be a rare, nearby analog of the ONC at a more advanced
stage of evolution. In this paper, we present a census of young stars, both
with and without disks, as well as protostars in the Cep OB3b cluster.  Using
this census, we show that the fraction of young stars with disks varies within the cluster.  Finally, we discuss reasons for this variation.

\section{OBSERVATIONS \& DATA ANALYSIS\label{obs}}

\subsection{Mid Infrared Data\label{ir_obs}}

Mid-infrared (MIR) observations were obtained with the {\it Spitzer}
Infrared Array Camera (IRAC) \citep{fazio04} and the Multi-band
Imaging Photometer System (MIPS) \citep{rieke04} as part of a {\it
  Spitzer} Cycle-2 GO survey (PID 20403) of the entire Cepheus OB3
molecular cloud.  The molecular cloud was mapped by IRAC at two
different epochs; the first on 2005 December 25 and the second 2006
August 8.  The map was further extended in two epochs (2007 August 17
and 2008 January 31) to cover the entire Cep~OB3b cluster as part of
the {\it Spitzer} program PID 40147.  Each field was observed four
times in 12 s High Dynamic Range (HDR) mode.  A total integration time
of 41.6 s was obtained at each wavelength for each field. 

The $\it{Spitzer}$/MIPS 24 $\mu$m images were obtained on 2007 February 25-26
and 2007 October 27 as part of the programs described above (PID 20403
and 40147), using scan maps at a medium scan rate with an offset of
$302''$ between scan legs.  Complete coverage was obtained only for
the 24~$\mu$m band; consequently, we only present results from this MIPS
band.  This paper focuses on a $49'$ $\times$ $33'$ region on the
north-west edge of the large Cep OB3 molecular cloud which encompasses Cep OB3b.

Mosaics of the IRAC data were constructed from the basic calibrated
data (BCD) products provided by the Spitzer Science Center using their
standard data processing pipeline versions S18.5 and 18.7.  Post-BCD
reduction of the {\it Spitzer}/IRAC data was executed using a software
package called ``Cluster Grinder'' \citep[see][]{gutermuth09}, developed
specifically to work on young clusters.  The package includes modules
for bright source artifact correction, cosmic ray removal, mosaicking,
and the visualization and photometry tool ``Photvis''
\citep{gutermuth08a} for point source extraction and
aperture photometry.  Radii of $2''.4$, $2''.4$, and $7''.2$ were used
for the aperture, inner, and outer sky annular limits, respectively
(choices of aperture and annuli are justified in \citet{allen08} and
\citet{gutermuth08a}).  

The instrumental magnitudes were derived directly from the BCD frames
converted into units of DN. For the 10.4 second frames, the adopted
zero-magnitudes were 22.00, 21.24, 19.04 and 19.43~mag for the
3.6, 4.5, 5.8 and 8~$\mu$m bands, respectively \citep{reach05}. These
included aperture corrections from a 2 pixel to 10 pixel aperture of
1.21, 1.23, 1.58, and 1.58, respectively, as tabulated in the IRAC
Data Handbook. The {\it Spitzer} band magnitude labels are denoted by
[wavelength in microns]: for example, $[3.6] = 15.16$ means that the
3.6 $\mu$m magnitude is 15.16. Ninety percent differential
completeness limits of the final mosaics are approximately $[3.6] =
15.16$, $[4.5] = 15.18$, $[5.8] = 13.71$, and $[8.0] = 11.72$
(see \citet{guter05} for the estimation technique).

\begin{figure}
\epsscale{.6}
\plotone{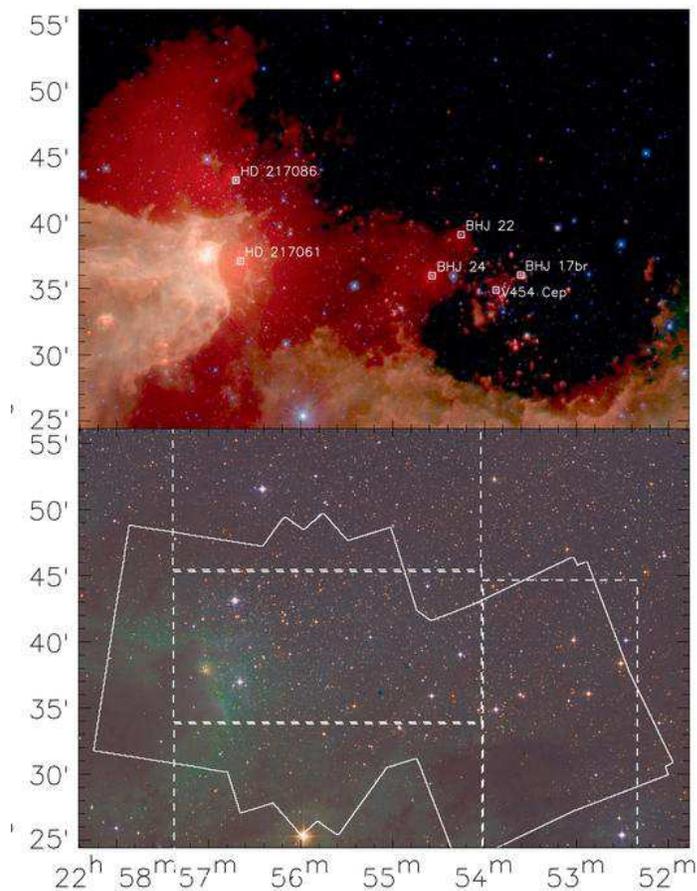}
\caption{Top: Three-color 3.6 (blue), 5.8 (green), and 24 $\mu$m (red)
  image of Cepheus OB3b region.  The extended pink emission is likely
  a mixture of emission from PAHs (particularly the 6.2 $\mu$m feature) mixed
  with thermal emission from dust grains in the 24 $\mu$m band.  This
  emission defines the border of a cavity in the molecular cloud
  within which the young cluster is located.  The extended red
  emission is due to warm dust at 24 $\mu$m in the HII region.  The
  massive stars in the region are labeled; the O7V star is HD 217086 and
  the early B stars are HD 217061, BHJ 22, BHJ 24, BHJ 17br and V454 Cep.  Bottom: Color composite Poss II image showing Cep OB3b in the visible.  The nebulous green emission is the HII region Sharpless 155.  The regions imaged by $\it{Chandra}$ for this study and from \citet{getman06}  are outlined with the solid white line.  The region imaged in the visible by \citet{mayne07} is outlined with the dashed white line.   \label{fig_mipsmap}}
\end{figure}

PID 20403 MIPS data were processed by the MIPS instrument teams' Data
Analysis Tool \citep{gordon05}, which calibrates the data, corrects
distortions, and rejects cosmic rays during the co-adding and
mosaicking of individual frames. The MIPS 24~$\mu$m mosaic created
from the four scan maps has a pixel size of $1''.245$ and units of DN
s$^{-1}$ pix$^{-1}$.  The MIPS field was extended in the PID 40147.
For this second observation, MIPS BCD data from the S16.1 pipeline were mosaicked using the
``Cluster Grinder'' package.  A pixel size of $2.547''$ was used for
this mosaic.  Point sources were identified in the {\it
  Spitzer}/MIPS data using Photvis; aperture photometry was then
obtained using $7''.4$, $14''.7$, and $24''.5$ for the aperture,
inner, and outer sky annular limits.  These magnitudes were then used as the initial guesses for point spread function (PSF) fitting photometry of the identified sources using the IDL implementation of DAOPHOT \citep{lands}; this version of DAOPHOT was modified to mask out saturated pixels in partially saturated PSFs \citep{Kryukova2012}.  Two point spread functions, one to fit unsaturated sources and a larger one to be used to fit partially saturated sources, were created from fifteen sample stars located across the MIPS field. The PSF generated for the unsaturated sources had a size of $62'' \times 62''$, while the PSF generated for the saturated sources was $87'' \times 87''$ in size. The fitting radii were  $2.5''$ and $12.45''$ for the unsaturated and saturated PSFs, respectively. The large fitting radius for the saturated PSF was needed to fit the wings of the PSF after the central, saturated pixels were excluded from the fit. The MIPS photometry were calibrated using a zero-point flux density of 7.17 Jy, referenced to Vega \citep{engel07}.  The ninety percent differential completeness limit is $[24] = 7.68$. A three-color rendering of the $\it{Spitzer}$ data can be seen in the top pane of Fig.~\ref{fig_mipsmap}.

\subsection{X-ray Data\label{xr_obs}}

X-ray data were obtained with the ACIS instrument aboard the $\it{Chandra}$ X-ray Observatory.  Two $17^{\prime} \times 17^{\prime}$ fields were imaged covering both sub-clusters.   Three images of the field covering part of the eastern sub-cluster were obtained with an aim point of 22:55:47.5, +62:38:10.3 (PIDs 9920, 10809 and 10810, with roll angles of $48^{\degree}$, $35^{\degree}$ and $72^{\degree}$ respectively), and  three images of the field covering the western sub-cluster were obtained with an aim point of 22:53:31.7, +62:35:33.4 (PIDs 9919, 10811 and 10812, with roll angles of $68^{\degree}$, $63^{\degree}$ and $63^{\degree}$ respectively).  Each image has an integration time of 28 ks.  The remainder of the eastern subcluster was previously imaged with $\it{Chandra}$ by \citet{getman06} (PID 3502, aim point of 22:56:49.4, +62:39:55.6 and roll angle of $7.9^{\degree}$).  The fields are outlined in the bottom pane of Fig.~\ref{fig_mipsmap}. 

These data have all been reduced in a similar manner using the $\it{Chandra}$ Interactive Analysis of Observations (CIAO) software tools.  The data were processed through the standard CIAO pipeline at the $\it{Chandra}$ X-ray Center, using their software version DS7.6.  This version of the pipeline automatically incorporated a noise correction for low energy events. This filter can remove good events  from the cores of bright point sources, resulting in an underestimation of the X-ray flux. In this case, the count rate did not exceed 0.02 counts per second for any source thus, event loss was not a concern. Background was nominal and non-variable.  

As the goal of the observations was the correlation of X-ray and infrared sources, point source detection was aggressive.  First the CIAO script fluximage was used.  The fluximage script creates exposure-corrected images of ACIS observations as well as the corresponding aspect and instrument maps which prepare the data for point source detection.  The script is especially efficient in cases in which multiple observations are used.  Sources were detected using the Wavdetect routine.  Wavdetect correlates the image with point spread functions (wavelets) of different scales and then searches the results for significant correlations.  In this case,  scales of 1 to 16 pixels were search on with logarithmic spacing.  Significances as low as 1 sigma were returned.  To ensure detection, we require a corresponding detection in the infrared to confirm the X-ray nature of the source.  As discussed in $\S$ \ref{xr_ind}, the vast majority of X-ray sources with visible and infrared counterparts are found toward the Cep OB3b isochrone.  THis shows that there are few spurious X-ray sources in our sample. 

\subsection{Near Infrared and Visible Data\label{ir_obs}}

Sources detected by {\it Spitzer} and $\it{Chandra}$ were matched with counterparts from the 2MASS
Point Source Catalog (PSC) and the  $BVI$ band catalog of \citet{mayne07}.  A matching radius of $2^{\prime \prime}$ was used in all cases.  The addition of these catalogs extends the wavelength coverage for these objects into the near-infrared ($\sim$1
$\mu$m) and the visible ($\sim$0.5$\mu$m).  A catalog combining the {\it Spitzer}, 2MASS and $\it{Chandra}$ data will be called the X-ray sample and a catalog combing the {\it Spitzer}, 2MASS and $VBI$ data will be called the visible sample.


\section{THE IDENTIFICATION OF YOUNG STELLAR OBJECTS  \label{identifying}}

\subsection{IR Identification of Young Stars with Circumstellar Material\label{ir_ind}}

\begin{figure}
\epsscale{.8}
\plotone{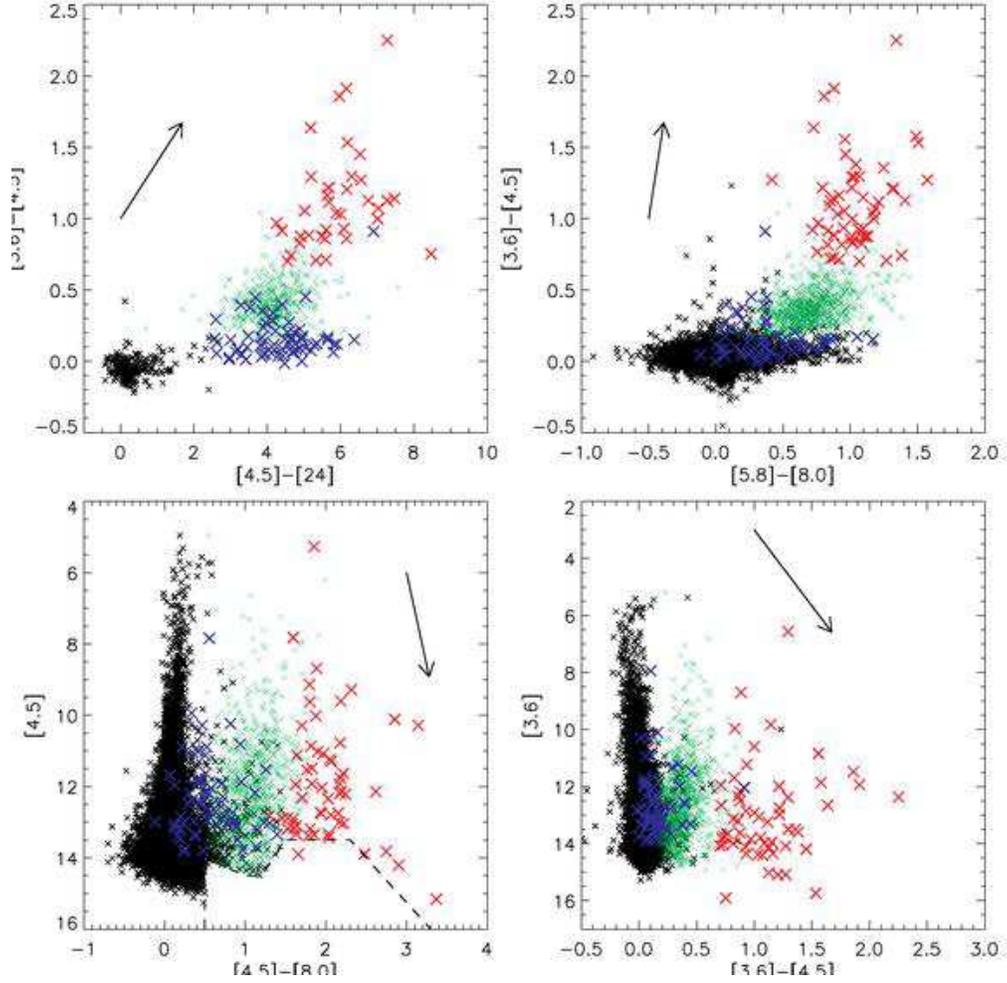}
\caption{IRAC/MIPS color-color and color-magnitude diagrams of Cep
  OB3b.  The colors of the markers correspond to the classifications
  determined with the three phases from Sec.~3. Red denotes Class I
  objects, green denotes Class II objects, blue denotes transition
  disk objects and black denotes Class III/field stars.  The dashed
  line in the [4.5]-[8] vs. [4.5] diagram (lower left plot) separates
  the region occupied by AGN contaminants.  All objects in this region
  were flagged as contaminants and deleted.  They were later included only if they have bright
  MIPS 24$\mu$m detection ($[24] < 7$) or very red IRAC/MIPS colors
  ($[3.6]-[5.8] > 0.5$ and $[4.5]-[24] > 4.5$ and $[8.0]-[24] > 4$).
  The black arrows show the reddening vectors for an
  $A_{k}=5$.  \label{cc_cm_fig}}

\end{figure}

\citet{gutermuth09} presented a new empirical three-phase scheme to
identify young stellar objects (YSOs) with infrared excess
(i.e. protostars with dusty infalling envelopes and young pre-main
sequence stars with dusty disks) which we adopt here. Those
classifications are illustrated in the IRAC/MIPS color-color and
color-magnitude diagrams shown in Fig.~\ref{cc_cm_fig}.

In phase 1, all four IRAC bands with photometric uncertainties $\sigma
< 0.2 $ mag are used to separate sources with infrared excesses due to
disks and infalling envelopes (Class II and Class 0/I, respectively) from pure
photospheres (Class III/field stars) based on $[3.6] - [5.8]$ and $[4.5] - [8]$ colors.  Various non-stellar contaminating sources
(e.g. PAH and AGN extra-galactic sources) also show infrared excesses;
they must be eliminated by several criteria described in detail by
\citet{gutermuth08a}, Appendix A. Galaxies with strong PAH emission
are eliminated by their colors.  AGNs mimic the colors of young
stellar objects and must be elimiminated by their faintness, as
determined by their location in the [4.5] vs. [4.5] - [8]
color-magnitude diagram (these cuts are shown in
Fig.~\ref{cc_cm_fig}).  Unresolved shocked emission blobs arising in
YSO outflows are identified by their strong 4.5 $\mu$m emission.  All
sources classified as contaminants were re-examined in the third phase of our identification process for excess
emission at 24 $\mu$m since some protostars might be masquerading as
contaminants. The remaining infrared excess sources which are not
removed as contaminants can be classified by [4.5] - [5.8] color as
Class 0/I objects (protostars) or Class II objects (pre-main sequence
star with disk).  The sources that do not show IR-excesses are
consistent with pure photospheric emssion between 1 and
8~$\mu$m. These include Class III (diskless pre-main sequence star),
field stars, as well as transition disk objects (stars with disks
exhibiting AU-scale inner holes) which will be
identified in later phases.  The distinct mid-IR colors of Class 0/I
objects, Class II objects and pure photospheres are clearly evident in
Fig.~\ref{cc_cm_fig}.

Phase 2 is applied to those sources which lack IRAC detections at 5.8
or 8.0 $\mu$m, but which have high quality  2MASS
detections ($\sigma < 0.1$) in at least $H$ and $K_s$ bands.  Line of sight extinction
to each of these sources is estimated using 2MASS colors (if $J, H,
K_s$ detections are all available) by de-reddening the sources that lie above the classical T-Tauri star (CTTS) locus of \citet{meyer97} or an adopted standard dwarf-star color \citep[$(J - H)_{0} = 0.6,$][]{bandb88} in $J - H$ and $H - K_s$ color space.  The sources that lie on or below these fiducial values are assumed to have no measurable reddening.  If high-quality $J$ detections
are unavailable, the line of sight extinction is estimated by dereddening the source to the CTTS locus in $H -
K_s$ and $[3.6] - [4.5]$ color space \citep[see][]{allen08} and using the reddening law of \citet{flaherty07}; this law was derived for star
forming regions similar to Cep OB3b.  Sources are then classified as either Class III/field objects, Class II objects or Class 0/1 objects based on their position in dereddened $(K - [3.6])_{0}$ and $([3.6] - [4.5])_{0}$ color space \citep{allen08}.  It must be noted that this phase applies primarily to lower-mass objects with disks, which are not detected at the longer IRAC wavelengths due to the increased background levels at those wavelengths.  Due to the rising spectral energy distribution (SED) of the protostars with envelopes, we expect most of them to be detected at the longer IRAC wavelengths and thus, we expect this phase to only add a few, if any, protostars to our catalog.  As expected, all the protostars in our catalog are classified during phase 1.

In addition, there are a handful of objects with Near-IR+IRAC colors
of Class IIIs, but which exhibit a 24 $\mu$m excess (e.g. $([5.8] -
[24]) >$ 1.5 and $([8] - [24]) >$ 1.5). These objects are classified
as transition disk objects, pre-main sequence stars surrounded by
disks that contain AU-scale inner holes \citep{muzerolle10}. Due to
the inner hole, these disks typically emit weakly at wavelengths $\le
8$~$\mu$m.

Phase 3 is applied to {\it all} sources classified in phase 1 with MIPS 24 $\mu$m detections
with $\sigma <$ 0.2 mag.  This phase re-examines all sources including contaminants and identifies Class 0/1 objects and transition disk objects on the basis of their 24 $\mu$m photometry.  In this phase, Class 0/I objects are defined as sources with $[5.8]-[24]) >$ 4.  The [3.6] - [4.5]
vs. [4.5] - [24] color-color diagram (Fig.~\ref{cc_cm_fig}) clearly shows the distinct colors
of our identified Class 0/I and Class II sources.

In total, there are 1135 objects in the Near-IR + IRAC dataset with evidence for dusty
envelopes or disks: these objects will be used in the following
analysis.  Of these, 56 are classified as protostars (Class 0/I), 1014
as pre-main sequence stars with disks (Class II), and 65 are
transition disk objects with AU-scale disk holes.  Tables \ref{classItable}, \ref{classIItable}, \ref{classtdtable} and \ref{classIIItable} list the sources identified as Class 0/I, Class II, transition disk and Class III, respectively.  A full source list
is included with the electronic version of this paper.


\subsection{X-ray Identification of Young Stars with Enhanced Coronal Activity\label{xr_ind}}

An estimate of the total number of young stars both with and without
disks requires a method for identifying and counting diskless Class III objects.  These pure 
photospheric sources cannot be distinguished from background field stars by their near to mid-IR 
colors.  In the following sub-sections, we use two separate methods to  quantify the number of diskless Class III objects.  
The first method directly identifies potential Class III objects from their X-ray flux density, and the second  
method distinguishes potential Class III objects using their visible magnitudes and colors.

Young stars have enhanced X-ray luminosities by orders of magnitude with respect to the main sequence (MS) in large part due to increased chromospheric activity \citep{feig09}.  By selecting objects that have both photospheric colors in the infrared as well as X-ray emission, we can identify diskless young members of the cluster.  To make this selection, we only need the spatial positions of the X-ray sources.  The analysis of the flux densities and other X-ray properties of these sources will be discussed in a future paper.  

We identify as X-ray members of Cep OB3b all sources that have a $\it{Chandra}$ X-ray detection, and have been classified as a Class 0/I/II/photospheric source using the $\it{Spitzer}$ and 2MASS three-phase classification scheme.  However, it is possible that coronal activity from another type of X-ray active field star, such as a flaring dMe star (the X-ray luminosity of dMe stars can reach up to log L$_{x} \sim 30$~erg s$^{-1}$ \citep{marino98}) could lead to a misidentification of a field star as a young star.  In comparison, the typical young star has an X-ray luminosity in the range of 28 $< $~log L$_{x} <$ 32 erg s$^{-1}$\citep{feig09}).  Thus, the X-ray active field stars may contaminate the $\it{Chandra}$ sample and be misclassified as a Class III source.  To estimate the potential magnitude of the background contamination, we examine the location of the X-ray sources in $V$ and $V - I$ color-magnitude space (right pane Fig.~\ref{fig_vis}).   We consider the sub-set of sources that have detections in all three data sets $\it{Spitzer}$, $\it{Chandra}$, and $BVI$.  This sample contains:  0 Class 0/I, 136 Class II, 9 transition disk and 329 Class III sources.  We assume that the potential  Class III sources that lie below the zero age main sequence (ZAMS) in Fig.~\ref{fig_vis} are all background contaminants (there are only two Class II sources in that region, they are the sources with  $V - I < 2$).  Further \citet{getman06} estimate that $13$ sources in their $17^{\prime} \times 17^{\prime}$ of Cep B are foreground stellar sources.  If we apply this number to out full $\it{Chandra}$ field of view, which is $729$ arcminute$^{2}$, we estimate that there will be 33 additional contaminants.  Combining these two estimates yield $11 \%$ (53/499) of the sources as contaminants.  Considering the full $\it{Chandra}$ field of view of $729$ arcminute$^{2}$, we have an X-ray contamination rate of 0.07 stars arcminute$^{2}$.

The X-ray sample includes: 5 Class 0/I, 233 Class II, 16 transition disk, and 499 Class III sources, of which 53 may be contaminating X-ray active field stars.  For reference, within the region imaged by $\it{Chandra}$ there are 36 Class 0/I, 810 Class II, 29 transition disk and 3213 photospheric sources (Class III or field stars) as classified via $\it{Spitzer}$ and 2MASS colors alone.  Thus, we are detecting $29\% (233/801)$ of the Class II sources and $55\%(16/29)$ of the transition disk sources with $\it{Chandra}$.

\subsection{Visible Wavelength Selection of Young Stars\label{df_vis}}
\begin{figure}
\epsscale{.8}
\plotone{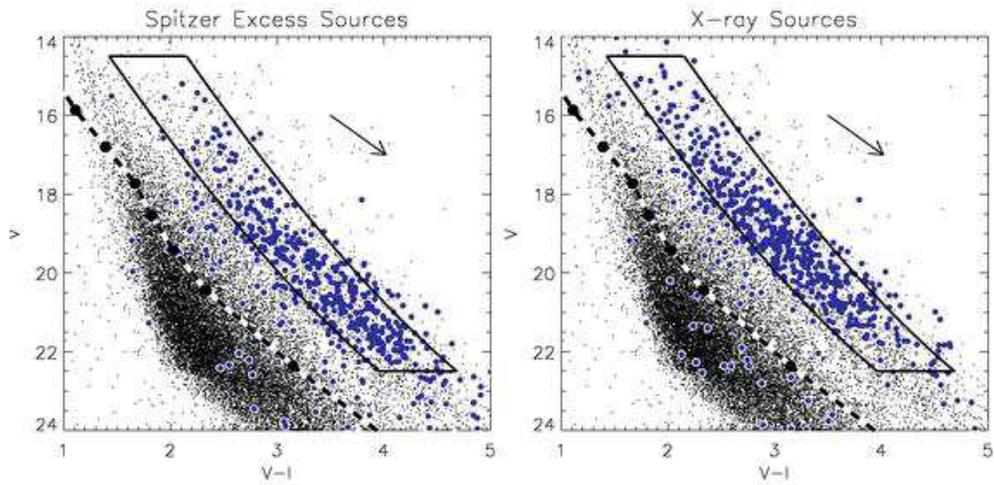}
\caption{$V$ vs. $V-I$ color-magnitude diagram of Cep OB3b. The black dots are sources in the Cep OB3b field with photometry that meets the photometric error threshold described in the text, and the blue dots are sources classified as pre-MS.  The region bounded by the black curve represents the YSO isochrone as described in the text.  The dashed white and black line are the ZAMS of \citet{siess00}, and the black arrow corresponds to 1 magnitude of extinction using the reddening law of \citet{rl85}.  The left plot shows sources classified based using $\it{Spitzer}$ colors and the right plot shows sources classified based on their detection in X-rays. \label{fig_vis}}
\end{figure}

We can also search for diskless young stars by utilizing the intrinsic magnitudes and colors of 
young stars to separate them from field stars in $V$ vs. $V-I$ space (Fig.~\ref{fig_vis}).  Young stars are still contracting to the MS, thus for a given 
temperature, young stars have a greater luminosity than field stars at an equal or greater distance.   The young stars occupy an isochrone (or bands of isochrones) that lies above the MS, and the extinction vector is approximately parallel to this isochrone.  Using 
the $V$ and $I$ band data of \citet{mayne07} for stars with $V$ and $V - I$ photometric errors less than $0.1$ mag, we empirically determine the YSO locus in $V$ and $V-I$ 
space.  
To determine the isochrone of young stars in the cluster, the locations of members identified by $\it{Spitzer}$ and $\it{Chandra}$ are plotted in $V$ and $V - I$ space (Fig.~\ref{fig_vis}).  The young stars are then divided into $0.75$ magnitude bins between $V$ magnitudes of 15 and 24, and a two step iterative process is used to determine the median $V - I$ color of the known YSO population in each bin.  First, the median color of the bin is calculated using all sources in the bin.  Next outliers greater than $3\sigma$ from the median are rejected and a new median and standard deviation are calculated.  The width of the YSO isochrone in color space 
is then taken to be between $-1.25\sigma$ and $2\sigma$ around the median color of the bin.  The median color of the 
YSO isochrone is fit with a third order polynomial, as are the outer bounds of the YSO isochrone; all stars between the two outer bounds are considered to be likely members.  The lower bound of the isochrone is cropped closer to the median value because the density of sources quickly increases in the direction perpendicular to the isochrone in the direction toward the ZAMS.  In particular, foreground dwarfs and background giants may masquerade as YSOs by appearing more luminous for a given color.  The outer bounds are shown in Fig~\ref{fig_vis}.   Hereafter, this selection of sources will be denoted by the visible sample.  

We can estimate an upper bound to the possible non-member stellar contamination (both foreground and background) by considering the $V$ and $I$ photometry from field 3a from \citet{pozzo03}, hereafter called the Pozzo field.  This is a $12^{\prime} \times 12^{\prime}$ field centered on RA: 22:50:21.14 and Dec: 62:46:44.88, on the north-west of the molecular clump Cep F.  We choose this region due to its location off the cluster and because it has the least number of $\it{Spitzer}$ identified excess sources; only four.  If we assume that these four sources are the only sources in this field that are members of Cep OB3b, then we can apply the same $V$ and $V - I$ selection criteria and consider all remaining sources that occupy the YSO dominated region of color-magnitude space as contaminants.  There are 98 sources that fit this criteria (102 sources in the YSO dominated region, minus the 4 known excess sources), for a contamination rate of $0.6$ stars arcminute$^{-2}$. 

Another way to estimate the number of non-member stellar contaminants in the $V$ and $V-I$ space is to extrapolate from the number of sources in the region of $V$ and $V-I$ space below the YSO isochrone into the YSO isochrone.  Assuming that there is a constant stellar density along the line of sight, then the number of stars of a given magnitude, $m$, is approximately $10^{0.6m}$ \citep{bm1998}.  We approximate the YSO isochrone with a linear strip in $V$ and $V-I$ space that is two magnitudes wide (bounded by the lines $V = 2.9(V - I) + 9.4$ and  $V = 2.9(V - I) + 11.4$), and consider the sources found in the one magnitude linear strip below the YSO isochrone,bounded by the lines $V = 2.9(V - I) + 11.4$,  $V = 2.9(V - I) + 12.4$, $V = 14.5$ and $V = 22.5$ to be non-member field stars.  Then the ratio of non-member field stars along the line of sight within the YSO isochrone to the number of stars within the field star isochrone is just 
\begin{equation}
\frac{N_{\mathrm{contaminants}}}{N_{\mathrm{field}}} = \frac{\int_{0}^{2} 10^{0.6m} \mathrm{d}m}{\int_{2}^{3} 10^{0.6m} \mathrm{d}m} = 0.314
\end{equation}
Since there are 1211 sources in the one magnitude field star strip, we estimate 381 contaminants in the YSO isochrone.  Dividing by the area covered by the visible data $(4\times 11^{\prime}\times 22^{\prime})$ leads to a contamination rate of  $0.4$ stars arcminute$^{-2}$.  This figure is similar to, but slightly lower than, the density of contamination derived from the Pozzo field.  Because the Pozzo field may contain some young sources due to its proximity to the molecular cloud, we adopt $0.4$ stars arcminute$^{-2}$ as the visible selection contamination rate.
 
The visible sample includes:  402 Class II, 23 transition disk, and 1270 Class III sources, of which 381 may be contaminants.  If we restrict the sample to only the sources that lie within the region with a field of view in common with the $\it{Chandra}$ images (we will call this region the overlap region), we find 372 Class II, 17 transition disk, and 914 Class III sources, with an estimated 232 contaminants.

\section{THE SIZE AND STRUCTURE OF CEP OB3B\label{distribution}}
\begin{figure}
\epsscale{.6}
\plotone{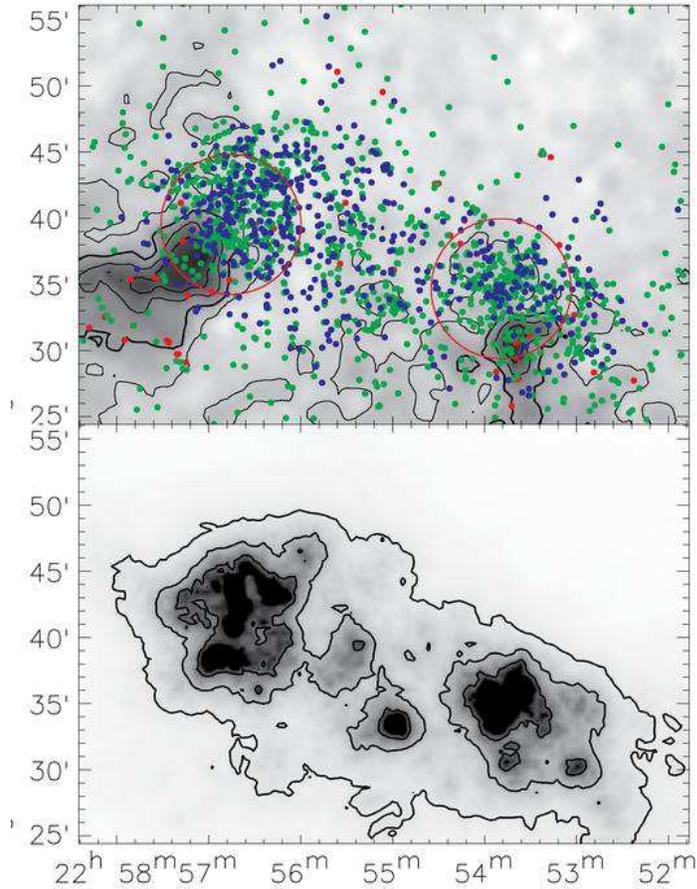}
\caption{Top: grayscale extinction map of same region with countours
  corresponding to $A_{V}$ of 2.5, 5.0, 7.5, and 10.  The red and green dots
  are $\it{Spitzer}$ classified young stars with circumstellar excess indicative of an envelope and a disk, respectively, and the blue dots
  are $\it{Chandra}$ detected young stars with increased coronal activity.  The large red circles show the adopted boundaries of the sub-clusters.  Bottom:  YSO surface density map created using IR, X-ray and visible selected YSO candidates.  YSO surface density was calculated using the 11 nearest neighbors.  Contours correspond to YSO surface densities of 10, 30 and 70 stars pc$^{-2}$. \label{fig_extmap}}
\end{figure}

The entire population of identified young stars (see
Fig.~\ref{fig_extmap}a) is found in a $\sim$ 7pc $\times$
10pc region.  The Class I and II objects are concentrated in two sub-clusters (hereafter, the eastern and western sub-clusters),
surrounded by an extended halo of these objects.  Furthermore, the two
sub-clusters are located primarily in a cavity with surfaces defined
by bright PAH emission; this cavity was probably evacuated by the OB
stars in the cluster (see Fig.~\ref{fig_mipsmap}a).  The eastern
sub-cluster has a Class II surface density peaking at about 260
pc$^{-2}$ and the western sub-cluster has a peak Class II surface density of about 310 pc$^{-2}$ (see Fig.~\ref{fig_extmap}b).  The sub-clusters are separated and surrounded by a diffuse 'halo' of young stars.  We define the halo as the region with a Class II surface density above 10~pc$^{-2}$.  The total area of the cluster, including the diffuse halo is about 70~pc$^{2}$.  There are 837 Class II sources within this area with an average surface density of about 12~pc$^{-2}$.

If we include the Class III YSOs determined via {\it Spitzer} and X-ray detections, the peak YSO densities reach $\sim$ 340 pc$^{-2}$ and $\sim$ 680 pc$^{-2}$ for the eastern and western sub-clusters, respectively.  This sample has 1317 sources within the 70~pc$^{2}$ region defined above as the total area of the cluster, for an average surface density, including the halo, of about 19~pc$^{-2}$.  If we consider the X-ray contamination rate of  0.07 stars arcminute$^{2}$ these surface densities will be reduced by about 10~pc$^{-2}$. 

If we consider the $\it{Spitzer}$ plus visible sample, the peak surface densities are  $\sim$ 540 pc$^{-2}$ and $\sim$ 730 pc$^{-2}$ for the eastern and western sub-clusters, respectively, and the average surface density including the halo is $\sim$ 30~pc$^{-2}$ (2112 sources / 70~pc$^{2}$).  Including the visible sample contamination rate of $0.4$ stars arcminute$^{-2}$, these values will be reduced by about 50~pc$^{-2}$.

An extinction map (see Fig.~\ref{fig_extmap}, top pane) was derived using
the $H - K_s$ colors of background stars from the 2MASS (Two-Micron
All Sky Survey) Point Source Catalog \citep{skrutskie06}.  The method
utilized is outlined in detail in \S3.1 of \citet{guter05} and
incorporates elements of the NICE \citep{lada94} and NICER algorithms
\citep{la2001}. Briefly, the line-of-sight extinction to each point on
a 31{\arcsec} grid is estimated by evaluating the mean and standard
deviation of the $H - K_s$ colors of the 20 nearest stars to the grid
point, and an iterative outlier $H - K_s$ color algorithm was used to
reject colors more than 3$\sigma$ from the mean until the mean
converged.   The extinction map was generated assuming an average $H - K_s$ color of 0.2 for background
stars, $A_{K_s} = 1.82(E_{H-Ks})$, and $A_V= 9.9A_{K_s}$ \citep{rl85}.  Typical resolution is $2'$, and
the range of A$_{V}$ is 0 - 12.6 mag in this map.  

The sub-clusters are associated with two distinct molecular clumps:
Cep B to the east and Cep F to the south-west using the nomenclature
of \citet{sargent77}. As noted, a large percentage of the pre-MS
population is in the low extinction cavity: $89\% (965/1079)$ of Class II and transition disk objects are projected
on a region with an $A_{V}<$ 5 based on the derived extinction map.  This observed configuration is
different from that found in the ONC. While the ONC is in front of and
partially embedded in the molecular cloud containing BN/KL, Cep~OB3b
lies mostly adjacent to two active molecular clumps.  This is advantageous because the detectability of low-mass pre-main sequence stars in Cep~OB3b by $\it{Spitzer}$ will be less affected by extinction and nebulosity than in the ONC.

The 56 protostars (i.e. sources classified as Class 0/I) are primarily
found in the vicinity of the Cep B and Cep F molecular clumps.   About
$41\% (23/56)$ are associated with high extinction, $A_{V}>$ 5, based on the derived extinction map.  Tis shows that star formation is continuing in these clumps.  The remaining 33 protostars
apparently are not associated with high extinction, although this may
be a short-coming in our extinction map because of its low spatial
resolution, and the consequent bias against detecting compact high
extinction structures.  On the other hand, some of these protostars
may be edge-on disks; \citet{gutermuth09} estimated that the number
of edge-on disks is at most 3.5\% of the total number of Class~II
objects. In Cep~OB3b this upper limit corresponds to $\sim$35 edge-on
disk systems. Since this number is comparable to the number of
protostars in Cep OB3b that are not associated with high extinction,
most of these YSOs classified as protostars may be edge-on disk systems.

\section{THE DISK FRACTION IN CEP OB3b AND ITS SUB-CLUSTERS\label{fractions}} 

\subsection{Disk Fraction in the Sub-Clusters\label{sub_fractions}} 


To estimate the disk fraction we require a census of the cluster members with and without disks.  While the number of members with disks is reliably determined from our sample of IR excess sources, to determine the number of diskless members we require a method to distinguish between diskless Class III and field stars seen in projection toward the cluster.    This has often been done by using measurements of the source density in off-cluster regions to estimate the number of field stars (see for instance \citet{carpenter00} and \citet{allen07}).  However, these methods require assumptions about the homogeneity of the background distribution of stars and line of sight extinction that may not hold in a large,  complex OB association with multiple epochs of star formation.  Thus, we use the direct identification of Class III members using either their X-ray flux or their position on the $V$ vs. $V - I$ diagram, as described in section~\ref{df_vis}. 

\begin{figure}
\epsscale{.8}
\plotone{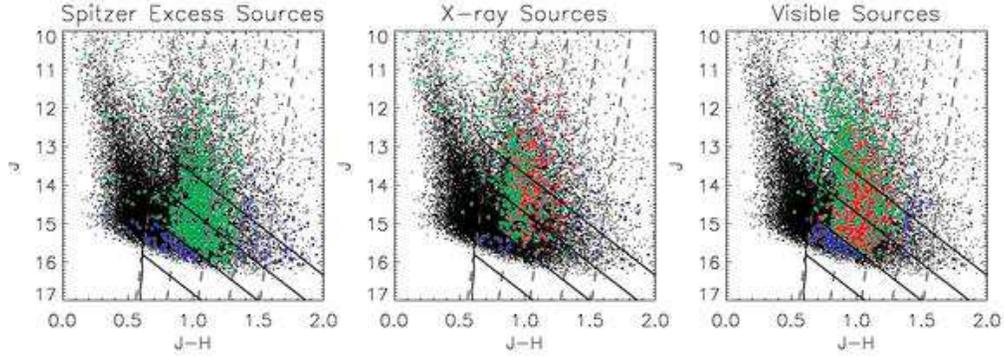}
\caption{  $J$ vs. $J-H$ color-magnitude diagram of Cep OB3b.  The black dots are sources in the Cep OB3b field with photometry that meets the photometric error threshold described in the text, while the green, red and blue dots are further selected as young sources based on $\it{Spitzer}$, $\it{Chandra}$ and/or visible photometry.  The red and green sources are excess and non-excess sources, respectively, that satisfy the magnitude-extinction requirements for the disk fraction analysis in section~\ref{fractions}, while the blue sources do not.  The upward sloping dashed lines represent $A_{V}$ values of 0, 2, 4, 6, and 8 magnitudes.  The downward sloping solid lines are reddening vectors for a $m_{J}$ = 13.01, 13.94, 14.75, and 15.81 star.  These magnitudes correspond to a 0.7, 0.35, 0.2 and 0.08 $\msol$ star at an age of 3 Myr at the adopted distance of 700 pc using the model of \citet{baraffe98}.  Each pane shows a different selection method.  From left to right: $\it{Spitzer}$ infrared excess sources, $\it{Chandra}$ X-ray sources and visible wavelength selected sources.  The criteria for each selection method are described in the text. \label{fig_nir}}
\end{figure}

Since disk sources are selected because of their increased flux in the infrared, sources with disks can be detected to fainter magnitudes, and higher extinctions, than purely photospheric sources.  To reduce the impact of this bias, we apply a near-IR magnitude and color requirement based on the position of the source on the  $J$ vs. $J - H$ color-magnitude diagram (Fig.~\ref{fig_nir}).  The sources that remain are those that are brighter than a de-reddened $J$-band magnitude of 14.75.  This magnitude and reddening vector corresponds to a 0.2 $\msol$ star at a distance of 700 pc using the models of  \citet{baraffe98}.  Further, these sources must have colors that indicate an extinction of less than $6~A_{V}$.  These limits were set based on a visual inspection of the positions of the sources in Fig.~\ref{fig_nir}.


After applying these magnitude and extinction cuts, we are left with 51 Class 0/I, 727 Class II and 56 transition disk sources from the $\it{Spitzer}$ and 2MASS sample.  For the X-ray sample 2 Class 0/I, 207 Class II, 16 transition disk and 451 Class III sources remain, and for the visible sample 345 Class II, 22 transition disk, and 1139  Class III sources remain.

We calculate the disk fraction for the entire cluster and for two $5'$ radius regions centered on the eastern (RA: 22:56:49.44, Dec: 62:39:55.54) and western (RA: 22:53:50.83, Dec: 62:35:36.37) sub-clusters, shown as red circles in Fig.~\ref{fig_extmap}a.  The disk fraction for the entire X-ray sample is $33\% \pm 2\%$ (207 Class II, 16 transition disk and 451 Class III).  The eastern sub-cluster has a disk fraction of $32\% \pm 4\%$ (62 Class II, 2 transition disk and 137 Class III), while for the western sub-cluster, the disk fraction is $50\% \pm 6\%$ (64 Class II, 7 transition disk and 71 Class III).  The disk fraction for the visible sample is $24\% \pm 1\%$ (345 Class II, 22 transition disk and 1139 Class III).  This is significantly lower that the disk fraction estimated using the X-ray sample.  The visible data used in this analysis, however, extend to the north of the region imaged by $\it{Chandra}$.  If we restrict our estimate to only those sources in the visible sample located in the region imaged by $\it{Chandra}$, the disk fraction is $29\% \pm 2\%$ (321 Class II, 17 transition disk and 811 Class III).  This is consistent with the disk fraction estimated with the X-ray sample.   Further, we find disk fractions of $34\% \pm 3\%$ -  (i.e., 103 Class II, 0 transition disk and 199 Class III) in the eastern sub-cluster and $44\% \pm 5\%$ -  (i.e., 71 Class II, 4 transition disk and 96 Class III) in the western sub-cluster.  Note that the fraction of sources with disks is similar regardless of the method used to identify the diskless Class III sources.  These results are summarized in Table~\ref{df_table}, where we tabulate the disk fractions with and without the $J$ vs. $J - H$ cuts.

Although we have minimized the contamination of the Class III sample by requiring either an X-ray detection or V vs V-I colors coincident with the CepOB3b isochrone, the residual contamination is the primary systematic uncertainty in our disk fraction measurement.  In Table~\ref{df_table}, we correct for the background contamination using the 0.4 arcminute$^{-2}$ contamination rate from $\S$ 3.  Since we do not know the $J$ and $H$-band magnitudes of the contamination, we do not apply the $J$ vs $J - H$ cut and background subtraction simultaneously. The correction for contamination raises the disk fraction because the background subtraction is only applied to the Class III objects.  For example, the background subtracted disk fractions for the overlap region increases to $36\% \pm 2\%$ for both the X-ray and visible sample.  Although the sub-clusters subtend a smaller region of the sky and hence have smaller rates of contamination; the subtraction of contamination can accentuate the difference in the disk fractions. In particular, if we consider the visible selection contamination rate of 0.4 stars arcminute$^{2}$, this is equivalent to 31 contaminants in each sub-cluster. After subtracting these from the Class III count, the disk fractions increase to $39\% \pm 4\%$ -  (i.e., 121 Class II, 0 transition disk and 191 Class III) in the eastern sub-cluster and  $53\% \pm 6\%$53 - ( i.e., 83 Class II, 4 transition disk and 75 Class III) in the western sub-cluster.  If we subtract out the potential contamination in the X-ray data (0.07 arcminute$^{-2}$) then this corresponds to a contamination rate of 6 contaminating sources in each sub cluster, resulting in disk fractions of $34\% \pm 4\%$ in the eastern sub-cluster  (72 class II, 2 transition disk and 145 class III) and $52\% \pm 6\%$ in the western sub-cluster (73 class II, 7 transition disk and 73 class III), see Table~\ref{df_table}.

We place the disk fraction estimates for Cep OB3b and its two sub-clusters within the context of other star forming regions by placing them on a plot similar to Figure 14 from \citet{hern08}, as seen in Fig.~\ref{fig_jesus}.  We adopt the disk fractions for the X-ray sample with the $J$ vs. $J - H$ cut ($32\% \pm 4\%$ in the eastern sub-cluster and $50\% \pm 6\%$ in the western sub-cluster); the systematic uncertainties due to Class III selection and contamination are similar to the formal uncertainties displayed in the figure.  The age is assumed to be between 3 and 5 Myr, represented by the horizontal uncertainty bars on the plot.

\subsection{Disk Fraction as a Function of Dereddened V-I Color\label{color_df}}

$\it{Spitzer}$ observations of young stellar clusters suggest that the disk fraction may be mass dependent, with the gas rich disks around more the massive stars disappearing first \citep{lada06, hern07, hern09}.  To search for a similar trend in our data, we use the dereddened $V - I$ color as a proxy for the mass, where sources are dereddened with the line of sight extinction estimated using the method described in $\S$~\ref{ir_ind} .  To directly compare the visible and X-ray samples, we consider those members detected in the $V$ and $I$-bands in the field of view common to both the visible and X-ray samples.  We bin the sources into three bins of blue, intermediate and red colors, with roughly 380 sources in each bin. The dereddened $V - I$ bounds for these three bins are: -0.5 to 1.05 mag, 1.05 to 1.72 mag and 1.72 to 4.35 mag, respectively.  We use the blue, intermediate and red bins as proxies for high, intermediate and low-mass young stars. To roughly estimate the corresponding mass ranges for to these color bins, we compare these dereddened colors to Table 5 in \citet{kenyon1995}.  These colors correspond to spectral type ranges of greater than K3, K3 to M0 and M0 to M6, respectively.   Using the tracks of \citet{siess00} and an age of 3 Myr, these correspond to masses of greater than 1.7~$\msol$, 1.7 to 0.6~$\msol$ and 0.6 to 0.2~$\msol$.  Table ~\ref{color_table} shows the results for the field of view common to the X-ray and visible light observations and for the sub-clusters, respectively. 

We would also like to estimate the potential number of non-member stellar contaminants in each of these color bins.  To do so we consider the same field star reference strip as defined in $\S$~\ref{df_vis}.  However, to determine the color bins above, we depend on the line of sight extinction estimates to each of the sources.  We do not know what these will be for the potential contaminants.  Instead we re-redden the bounds for the color bins using the median $A_{V}$ of all the sources considered to create the color bins.  The median $A_{V}$ of these sources is 3.3 mag, and the resulting $V - I$ color bounds are 1.22 to 2.77 mag, 2.77 to 3.44 mag and 3.44 to 6.07 mag for the blue, intermediate and red bins, respectively.  Since there are 739, 340 and 115 sources in the blue, intermediate and red sections of the field star strip, we estimate 232, 107 and 36 contaminants using Equation 1, and a contamination rate of 0.24, 0.11 and 0.04 stars arcminute$^{-2}$, respectively.

Table~\ref{color_table} shows the disk fraction as a function of dereddened $V - I$ color bin.  For the common field of view, the disk fraction is the same for all three bins of the X-ray sample, roughly $30 \%\pm 5\%$, whereas for the visible sample, the intermediate bin ($31 \%\pm 3\%$ for 1.7 to 0.6~$\msol$) has a slightly lower disk fraction than the reddest bin ($38 \%\pm 3\%$ for 0.6 to 0.2~$\msol$), and the blue bin ($19 \%\pm 2\%$ for $>$ 1.7~$\msol$) shows a substantially lower disk fraction than the intermediate bin.  Note that if we subtract the potential contaminants from the visible sample, the large discrepancy between the visible sample disk fraction ($30 \%\pm 4\%$) and the X-ray sample disk ($30 \%\pm 5\%$) in the blue bin disappears, while for the red bin the disk fractions differ by about 1.7 $\sigma$ ($41 \%\pm 3\%$ for the visible sample and $31 \%\pm 5\%$ for the X-ray sample).  Examination of the right pane of Figure~\ref{fig_vis} shows that the discrepancy in the red bin could be due to a lack of X-ray sensitivity to the reddest, lowest mass members. 

Looking at the sub-clusters in Table~\ref{color_table}, we find that the disk fraction does not vary greatly between the three bins for the western sub-cluster.  In contrast, the disk fraction in the blue bin ($19 \%\pm 4\%$ $>$ 1.7~$\msol$ ) for the eastern sub-cluster is half that of the disk fraction in the intermediate bin ($40 \%\pm 6\%$ 1.7 to 0.6~$\msol$). If we examine  the difference in the disk fraction between the two sub-clusters, this difference is greater that 2$\sigma$ for both the visible sample and the X-ray sample in the blue bin ($>$ 1.7 $\msol$).   The disk fraction can be found in Table~\ref{color_table}.  In contrast, for the intermediate bin (1.7-0.6 $\msol$) the disk fractions of the two sub-clusters are within 1$\sigma$ for the visible sample and 1.2 $\sigma$ for the X-ray sample.  For reddest bin (0.6 to 0.2 $\msol$), the difference in disk fractions  is about 1.3$\sigma$ for the X-ray sample and 1.2$\sigma$ for the visible sample.  Subtracting the potential contaminants from the visible sample increases the difference in the disk fraction between the two sub-clusters in the bluest bin.  The intermediate bin still has no difference, and the red bin has a difference of 1.1$\sigma$.  Thus, the lowest disk fractions and the largest difference in the disk fraction between the sub-clusters is in the bluest $V - I$ bin with the highest mass stars.  As we will discuss in $\S$~\ref{ages}, the observed difference for the bluest stars may be due to a systematically older age in the eastern sub-cluster.

Broad-band magnitude, and color, determinations of stellar mass are highly uncertain due to the number of non-mass related factors that could affect the magnitude, and color, of a young source, such as age, accretion, etc.  It is therefore prudent to base mass estimates of young stars from more robust measurements such as their spectral type, which will be done in a future paper.  We defer further investigations of the stellar mass dependence of the disk fraction until this paper.


\subsection{Spatial Variations in the Disk Fraction\label{spatial}}

The different disk fractions found in the two sub-clusters motivates an examination of the spatial dependence of the disk fraction.  We use two methods for visualizing the spatial variation of the disk fraction in Cep OB3b.  First, we create maps of the disk fraction throughout the cluster.  We created two disk fraction maps utilizing the X-ray sample and the visible sample.  To make these maps we first created a grid covering the entire region with nodes spaced every $\sim 50^{\prime\prime}$.  The disk fraction at each node was then determined using all classified sources within a $2^{\prime}$ radius of that node.  A map was then created using the disk fraction at a given node as the value for the respective ``pixel" on the map.  Fig.~\ref{fig_dfmap} shows the results of this process for both samples, and Fig.~\ref{fig_df_comp} shows a `pixel by pixel' comparison of the two maps.  Secondly, we consider the disk fraction in annular regions in the eastern sub-cluster centered on the O7V star HD271086 or the western sub-cluster centered on the B1V star V454 Cep, as seen in Fig.~\ref{fig_xr_distro} for the X-ray selected set and Fig.~\ref{fig_vis_distro} for the visible selected set.  Note that since the background contamination may be spatially variable due to spatially varying extinction, we do not correct for contamination.  Correcting the Class III sample for contamination would systematically increase the disk fraction and the amplitude of the observed variations.  

Both methods of visualizing the spatial variation in the disk fraction show the same general morphology.  The eastern sub-cluster is characterized by a broad area of fairly homogenous disk fraction, and as shown in a recent paper by \citet{getman09}, the disk fraction increases from about $30\%$ to over $50\%$ as one traces a path from the O star, to the Cep B molecular clump and its embedded population (Fig.~\ref{fig_dfmap}).  The western sub-cluster, however, is characterized by a high disk fraction in the center of the sub-cluster that drops off radially outward.  Note that the visible sample is biased toward the less embedded sources, and the map derived with this sample does not find some of the smaller clumps of stars with high disk fractions apparent in the X-ray map.  

One question we must ask is whether the observed variation in the disk fraction is consistent with random fluctuations.  To address this question, we focus on the different disk fractions observed toward the two sub-clusters.  To consider the range of disk fraction ratios we would expect due to statistical fluctuations, we perform a Monte-Carlo simulation.  We assume that each member of the cluster has a probability of harboring a disk to be equal to the disk fraction of the $\it{entire}$ region.  For each source in a given sub-cluster we draw a random number from a uniform distribution between zero and one using the IDL routine  RANDOMU.  If the random value associated with a given source has a value less than or equal to the disk fraction for the $\it{entire}$ sample it is considered to be a disk source, otherwise it is considered to be a non-disk source.  Once this is done for each source, the disk fractions for the sub-clusters are calculated, and their ratio is determined for this particular realization.  This is then computed for 10,000 realizations.   Fig.~\ref{fig_mcrx} shows the distribution of ratios derived using the visible sample, in the right pane, and the X-ray sample, in the left pane.  The actual ratio is given by the red line.  This exercise shows that the disk fractions in the two sub-cluster are different with a $95\%$ confidence. 

\subsection{Estimate of Total Cluster Membership\label{total_member}}

We can use the disk fractions measured in the sub-clusters to estimate
the total number of cluster members from the number of IR-excess
sources.  The $\it{Spitzer}$ data shows 1135 infrared excess sources, the vast majority of which are young stars with disks.  From Fig.~\ref{fig_nir}, we
ascertain that most of the sources above 0.2 $\msol$ are detected (adopting the \citet{baraffe98} tracks for a 3 Myr stars).   Using the average disk fraction determined from the X-ray sample, $33\% \pm 2\%$, we then estimate that there are about 3000 total members in the Cep OB3b cluster down to the approximate mass limit of our survey.

\section{DISCUSSION \label{discussion}} 

\subsection{Comparison of Cep OB3b to Other Young Clusters  \label{comparison}} 

Our estimated membership of Cep OB3b of 3000 primarily low mass stars, shows that
Cep~OB3b is comparable in size and membership to the ONC \citep[$\sim 1000$ stars in a 10pc $\times$ 5pc
  region]{allen07,carpenter00}, although the peak surface densities
are considerably less dense than the peak densities in the ONC (\citet{hill98}; Megeath et al. (in preparation)).  The content of massive stars in
Cep~OB3b is only slightly lower than that of the ONC: one O7 star in Cep OB3b compared to an
O6 and an O9.5 in the ONC; \citet{muench10}.  These results suggest
that Cep~OB3b is the second nearest, large, ($>1000$ star) young ($<$
5 Myr) stellar cluster to the Sun.  Furthermore, the published isochrones and the
relatively low extinction toward the region suggest that Cep~OB3b is
older and more evolved than the ONC.  The older ages and lower stellar ages suggest that this cluster has expanded since it dispersed its parental gas.

Other than the ONC, there are no known clusters within 1~kpc that are
comparable to Cep OB3b in age,  size and average density.  The largest known young cluster 
next to the ONC and CepOB3b within 1 kpc is in the NGC 2264 region at 760
pc, part of the Mon OB1 molecular cloud. This cluster contains $\sim
450$ young stars with disks, less than half as many as CepO3b \citep{sung09}.  The next largest known young clusters within 1~kpc, ranked by number of known disk sources, are:     IC 348, $\sim$400 \citep{muench07}; Sigma Ori, 336 \citep{hern07}; TR 37, $\sim$200 \citep{sicilia06}; Mon R2, 235 \citep{gutermuth09}; IRAS 20050, 177 \citep{gutermuth09}; AFGL 490, 161 \citep{gutermuth09}; IC5146, 149 \citep{gutermuth09}.  A major uncertainty in comparing the sizes of clusters is the number of diskless pre-main sequence stars.  Since the membership counts listed in this section are primarily from $\it{Spitzer}$ counts of disk sources, they are lower limits to the actual membership counts.  Nevertheless, the number of known disk sources in these other clusters are much less than the number of known disk sources in Cep OB3b.

\subsection{Disk Ablation from UV Radiation Field  \label{ablation}} 

It has been known for over a decade that UV radiation
from O-stars can ablate the outer regions of disks, a process that has
been directly observed in the Orion Nebula Cluster
\citep{bally98}. The IR-excesses detected by $\it{Spitzer}$ originate from
the warm inner disk, while photoevaporation primarily truncates the
outer disk \citep{adams04}.  However, disks may be rapidly destroyed
by a combination of photoevaporation and viscous accretion
\citep{matsu03}.  Evidence for the destruction of inner and outer
disks has been found in previous $\it{Spitzer}$ observations of young
clusters. In NGC 2244, the Rosette Nebula, at a distance of 1.6 kpc,
\citet{balog06} find outer disks being ablated near a massive O star.
\citet{balog07} also find evidence that the fraction of young stars
with disks drops near the massive O stars, evidence for the
destruction of disks by the O stars.  Similarly, \citet{sung09}
examine the young cluster NGC 2264 and find that the disk fraction
tends to increase radially away from the O7V star S Mon.

The only known O star within the Cep~OB3b cluster is the O7Vn star HD 217086
\citep{herrero92}, which is the primary ionizing source for the HII
region Sh 155, and is illuminating and photoevaporating the surface of
the nearby cloud as delineated by bright nebulosity in the IRAC and
MIPS bands. HD 217086 is also situated close to the highest projected
surface density of YSOs in the eastern sub-cluster
(Fig.~\ref{fig_extmap}b).  A B1V star is also coincident with this
sub-cluster.  In contrast, a B0V and three B1V stars are found in the
western sub-cluster.  The Lyman continuum production rate of the O7
star is nine times higher than that of a B0V star \citep{vacca96}, thus we must ask if the difference in the observed disk fractions in the two sub-clusters is due to a much higher incident EUV flux in the eastern sub-cluster \citep{adams04}. 

A signature of disk photoevaporation would be a decrease in the disk fraction radially toward the O star.  Figure \ref{fig_xr_distro} shows the disk fraction as a function of radial distance away from a massive star in each sub-cluster, using the X-ray sample.  The top pane is centered on the O7V star HD 217086, in the eastern sub-cluster.  Each annulus centered on this star has a width of $2^{\prime}$ within which the disk fraction is estimated by counting the number of sources with and without a $\it{Spitzer}$ excess.  The bottom pane is centered on the B1V star HD 216711 (V 454 Cep), in the western sub-cluster, and the disk fraction is estimated in the same manner as the top pane.  The disk fraction is fairly constant as a function of radial distance from the O7V star in the eastern sub-cluster, while it increases significantly toward the B1V star in the western sub-cluster.  Figure \ref{fig_vis_distro} shows the same radial disk fraction distribution if the visible sample is used to select sources.  We find no evidence for disk photoevaporation in these plots.

To consider whether mixing could influence the radial disk fraction distribution, i.e. if stars move into and away from the harsh radiation environment near the O7V star on timescales short enough to erase any disk fraction differences due to photoevaporation, we use the virial theorem to estimate the velocity of a typical star in the innermost $2^{\prime}$ radius region of each sub-cluster.  We assume that each star has the same mass ($0.5~\msol$), an isotropic velocity distribution, and experiences the same gravitational potential.  We assume the cluster is fully exposed, so that the mass of each sub-cluster is just the number of  sources times the assumed average mass and is estimated to be $46 \times 0.5~\msol$ = $23\msol$ for the western sub-cluster and  $75 \times 0.5~\msol$ = $37.5\msol$ for the eastern sub-cluster.  Note, here we consider $\it{all}$ sources and do not apply the magnitude and color cuts of section~\ref{fractions}.  The typical velocity then would be  $0.27$ km~$s^{-1}$ in the western sub-cluster and $0.22$ km~$s^{-1}$ in the eastern sub-cluster.  These correspond to a time of roughly 2 Myr to move $2^{\prime}$ in projection through the sub-clusters.  This time is less than the lower end of the estimated age range of 3-5 Myr.  Therefore, even though the radial disk fraction distribution near the O7V star does not indicate that photoevaporatation is the cause of the disk fraction difference between the two sub-clusters, the photoevaporation of disks in the eastern sub-cluster cannot be ruled out based on the radial disk fraction distribution alone. 

Finally, if disk photoevaporation is responsible for the different disk fractions in the two sub-clusters, we would expect the lowest mass stars to be affected the most due to their lower gravitational binding energy \citep{adams04}.  Instead, in Table~\ref{color_table}, we find the stars with the bluest V-I colors, and hence highest masses,  have the largest discrepancy in their disk fractions.  This is evidence against disk photoevaporation.

\subsection{The Ages of Sub-Clusters and Age Spreads \label{ages}}

\begin{figure}
\epsscale{.8}
\plotone{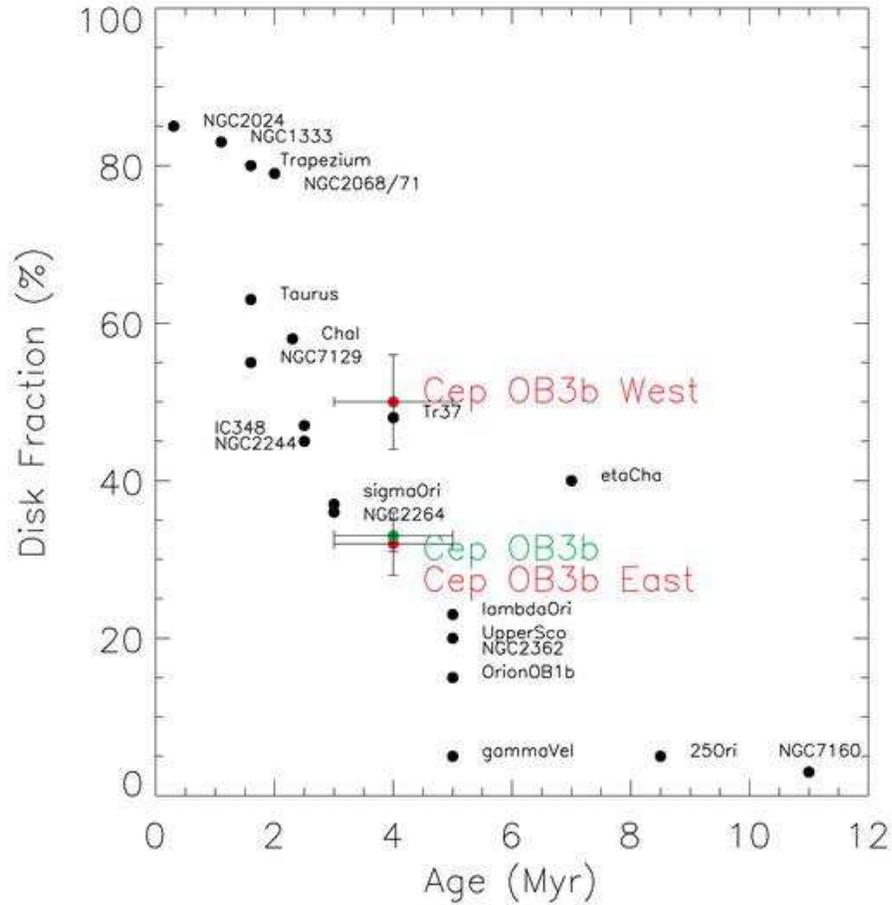}
\caption{Disk fraction vs. age for young stellar clusters in the literature.  This plot is taken from Figure 14 from \citet{hern08} and augmented with the disk fractions and assumed age of the Cep OB3b sub-clusters in red and the average disk fraction and assumed age of the entire Cep OB3b region in green.  \label{fig_jesus}}
\end{figure}

\begin{figure}
\epsscale{.6}
\plotone{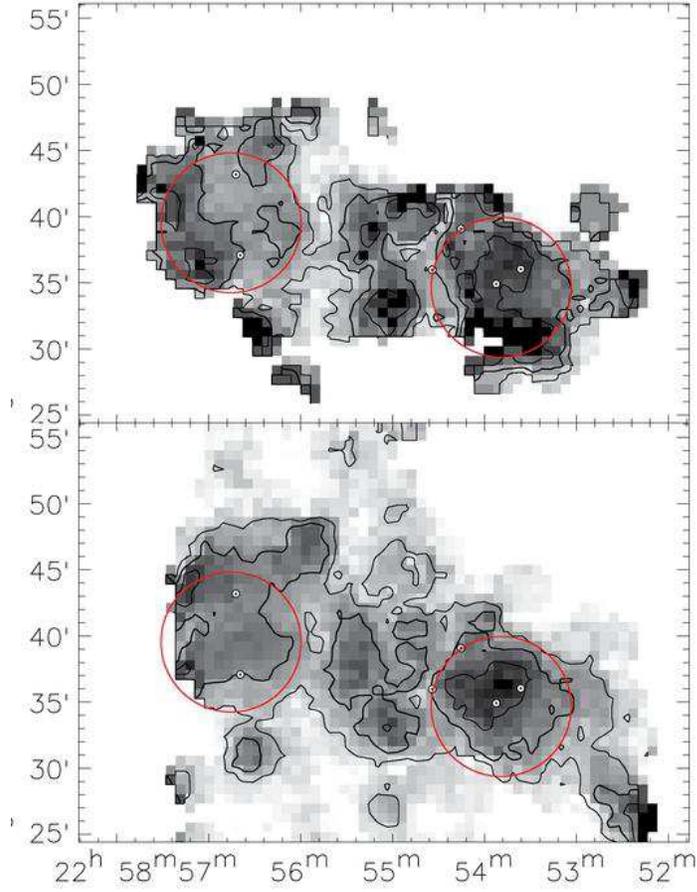}
\caption{Top: Disk fraction map created with the X-ray selected members.  Contours represent disk fractions of 20, 34, and 50 percent.  Red circles outline the sub-clusters and the white dots are the locations of the massive stars.  Bottom:  Disk fraction map created with the visible selected members.  Contours are at the same levels as above. \label{fig_dfmap}}
\end{figure}

\begin{figure}
\epsscale{.8}
\plotone{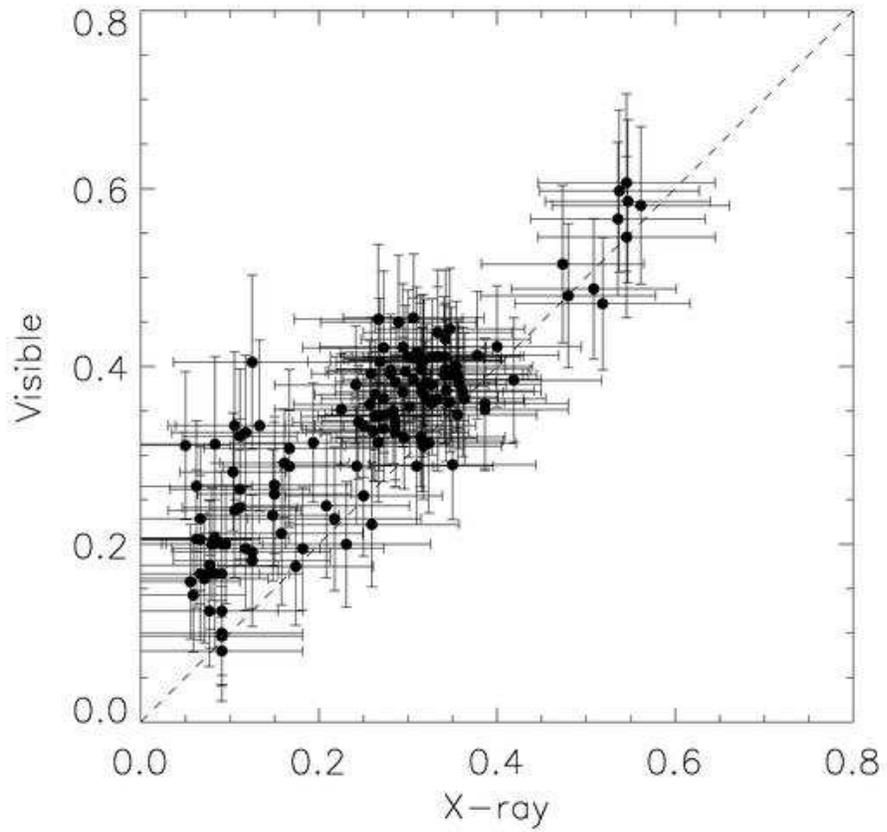}
\caption{A pixel by pixel comparison of the disk fraction maps created with $\it{Chandra}$ selected YSOs and visible selected YSOs.  The pixels considered are those with a measurement uncertainty of less than 10 percent.  The dashed line, with a slope of unity, is plotted for reference. \label{fig_df_comp}}
\end{figure}

\begin{figure}
\epsscale{.8}
\plotone{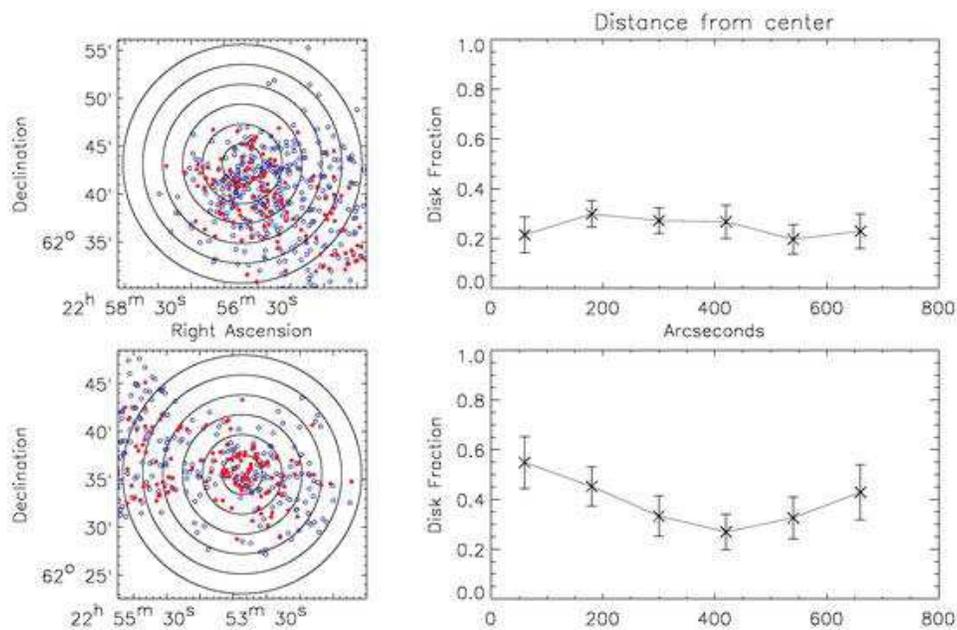}
\caption{Disk fraction as a function of radial position for the X-ray selected young stars with (filled red circles) and without (open blue circles) disks, subject to the mass and extinction requirements discussed in the text.  The top plots are centered on the O7V star HD 217086, while the bottom lots are centered on V454 Cep, a B1V star at the center of the western sub-cluster.  The disk fraction is determined within each annuli, which are separated by $2^{\prime}$.  Note, the disk fraction in the outer annulus of the western sub-cluster is increased by the small grouping of stars with disks between the two sub-clusters.    \label{fig_xr_distro}}
\end{figure}

\begin{figure}
\epsscale{.8}
\plotone{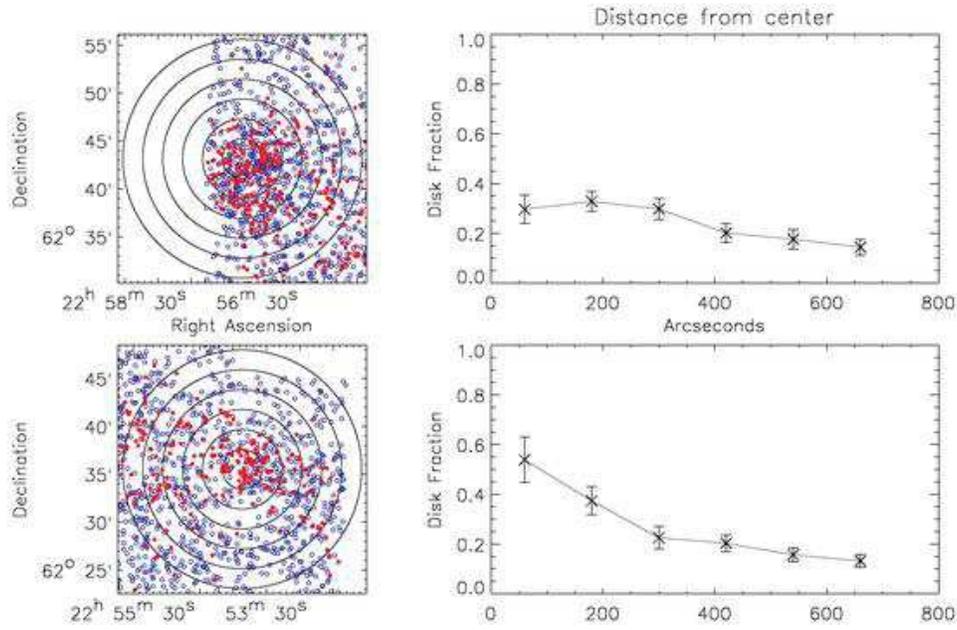}
\caption{Disk fraction as a function of radial position for the visible selected young stars with (filled red circles) and without (open blue circles) disks, subject to the mass and extinction requirements discussed in the text.  The center positions and annuli are the same as Figure~\ref{fig_xr_distro}.  Note, that the visible sample does not pick up the same increase in disk fraction in the outer annuli as the X-ray sample.   This could be due to preferentially selecting against reddened disk sources when requiring a $V$-band detection.  \label{fig_vis_distro}}
\end{figure}

\begin{figure}
\epsscale{.8}
\plottwo{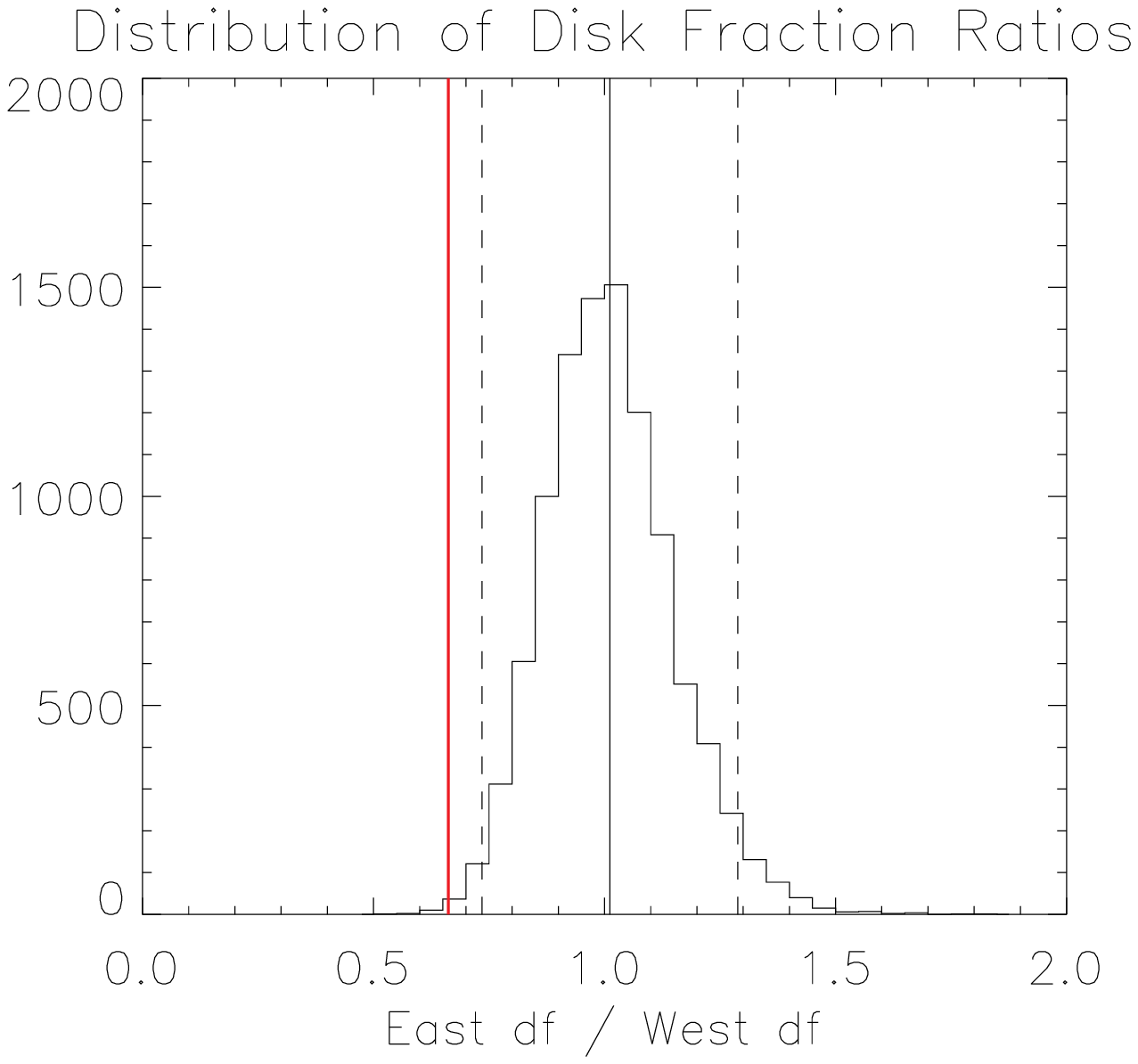}{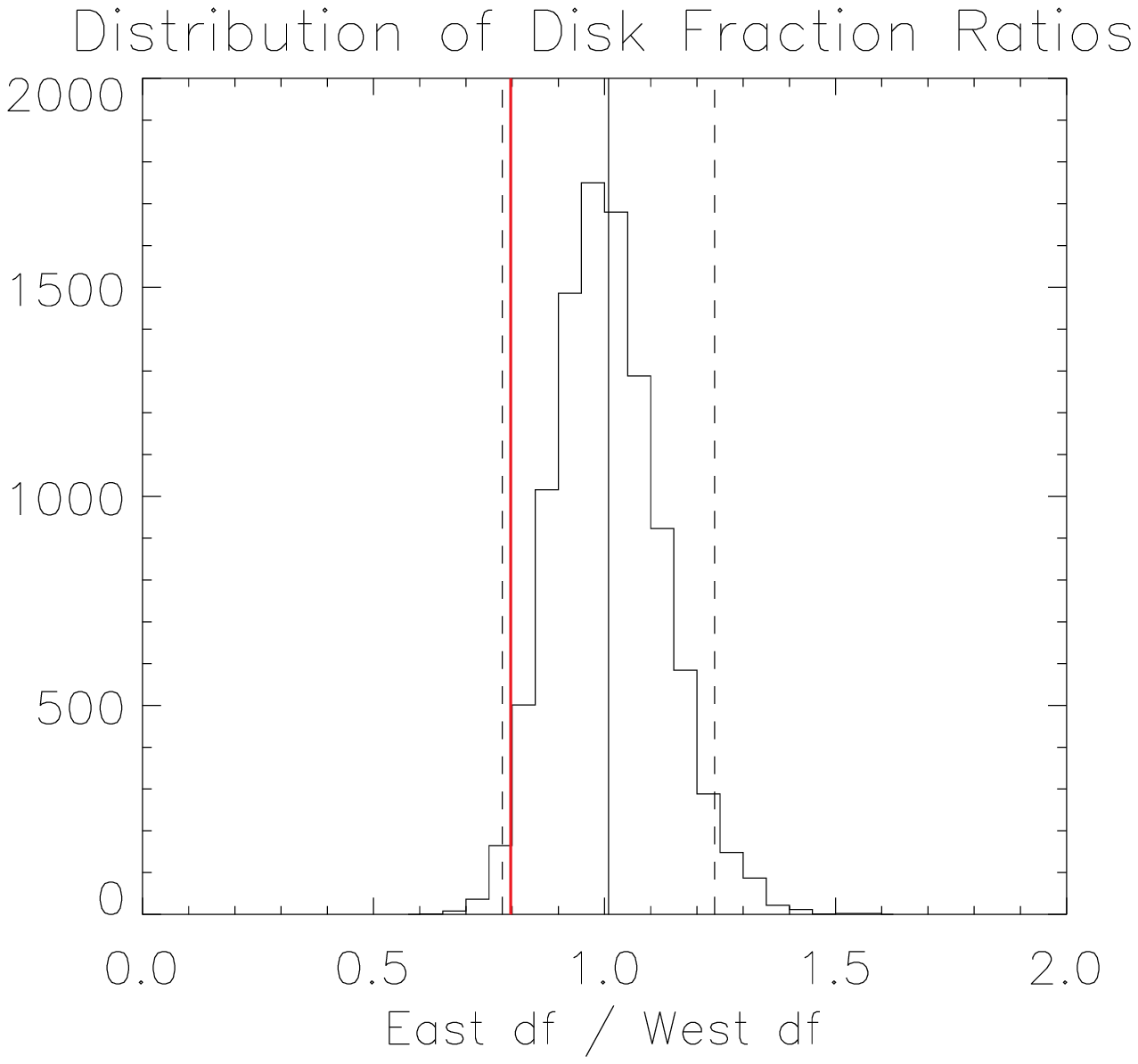}
\caption{Results of a Monte Carlo simulation of the disk fractions in the sub-clusters.  The histogram show the results for 10,000 realizations of the simulation, the solid black line is the mean, the dashed black lines are 2$\sigma$ from the mean and the red line is the actual ratio.  The left pane shows the X-ray selected sources, and the right pane shows the visible selected sources. \label{fig_mcrx}}
\end{figure}

Variations in the observed disk fraction may also be due to variations in the ages of the members.  For instance, \citet{hern08} show that the
disk fraction decreases from 80\% to 30\% as a stellar
cluster/association's age increases from 1 to 5 Myr (see Fig.~\ref{fig_jesus}).  An age variation due to an evolutionary gradient was proposed in \citet{getman09}.  Using deep IRAC observations of the eastern sub-cluster to supplement
their earlier X-ray observations, they have evaluated disk fractions
as a function of radius from the hot molecular core Cep B, which is in
the molecular clump at the edge of the eastern sub-cluster \citep{beuther00}.
They find disk fractions of $\sim75\%$ within
$3\arcmin$ from Cep~B, $\sim55\%$ within an annulus $3\arcmin$ to
$7\arcmin$ from Cep~B, and $\sim20\%$ within an annulus $9\arcmin$ to
$12\arcmin$ from Cep~B.  They interpret this spatial gradient as
temporal, with young stars dominant in the hot molecular core and
progressively older stars with increasing distance from the core.  The
temporal gradient results from a sequence of star formation triggered
by compression driven by the UV radiation from the O7V star.  Our
eastern sub-cluster includes most of the stars in these annuli, and we
examined the disk fraction for the X-ray sample of young
stars in two sections within the $5'$ region centered on the eastern
sub-cluster.  The first section is toward the high extinction region
of the cloud, and the second section is toward the lower extinction of
the S155 HII region and HD 217086.  We confirm that there is a higher
disk fraction toward the cloud than toward HD 217086, see $\S$ $\ref{spatial}$ and Figures $\ref{fig_extmap}$ and $\ref{fig_dfmap}$. 

\begin{figure}
\epsscale{.8}
\plotone{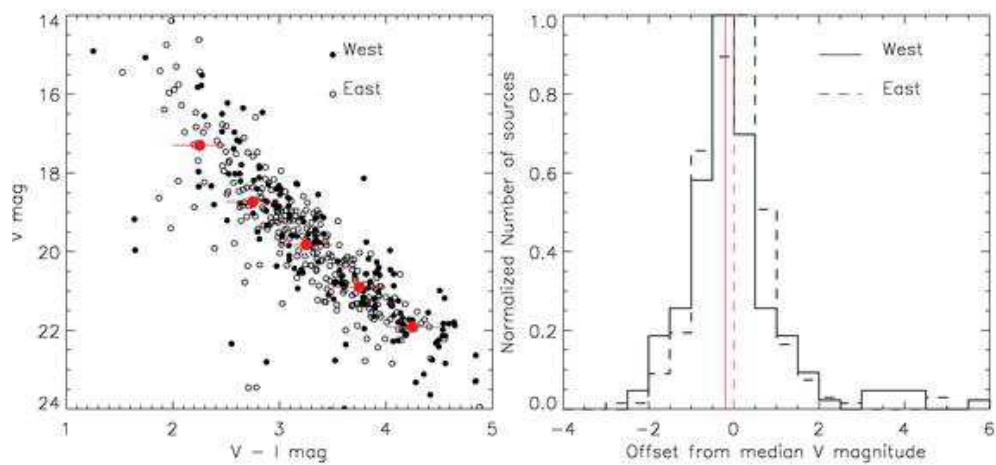}
\caption{Left: $V$ vs. $V-I$ color-magnitude diagram for the members of the eastern sub-cluster, open circles, and western sub-cluster, filled circles.  Filled red circles  are the median $V$ magnitudes for the color bins and the red lines show the width of each bin.  Right:  distribution of $V$ magnitude deviations from the median for the young sources in 0.5 magnitude color bins.  The dashed and solid red lines show the median color deviation for the eastern and western sub-clusters, respectively.      \label{v_vmi_sub}}
\end{figure}

To examine the possibility that an age difference could explain the difference in disk fraction between the two sub-clusters, we use the $V$ and $I$ data from \citet{mayne07} of the $\it{Spitzer}$ and $\it{Chandra}$ selected young sample.  We divide the sample into two sets, one for each of the two $5^{\prime}$ radius sub-cluster regions defined above.  Fig.~\ref{v_vmi_sub}a shows the location of these sources in $V$ vs. $V - I$ color-magnitude space.  Note, that the reddening vector in V vs. $V - I$ space is approximately parallel to the YSO locus, thus any extinction difference between the sub-clusters cannot masquerade as an age difference. 

By binning the $V - I$ axis into half magnitude bins between $V - I$ colors of $2$ and $4.5$, we find the median $V$ magnitude of all the young sources in each bin.  We then examine the offset of the $V$ magnitude distribution for each sub-cluster from the median $V$ magnitude by subtracting the median $V$ magnitude of a given color bin from the $V$ magnitude for each source in the same color bin.   Next, we plot the normalized histogram of the distribution of the offsets for each sub-cluster.  Due to the larger luminosity of younger stars of a given mass, we would expect a younger subset of stars to have a negative offset from the overall median.  Conversely, an older subset would have a positive offset.   An examination of Fig.~\ref{v_vmi_sub} shows that the sub-sample of sources from the western sub-cluster has a median offset that is slightly negative (-0.2 mag), while the eastern sub-cluster has an offset consistent with zero.  If we consider the $4$ Myr isochrone from the model of \citet{baraffe98} and shift it by an amount $0.2$ mag along the V magnitude axis in $V$ vs. $V - I$ color-magnitude space, this shift corresponds to an age difference of $1$ Myr for stars within the $0.2~\msol$ to $1.0~\msol$ mass range.

The full width half maximum of the distributions of offsets in Fig.~\ref{v_vmi_sub}  is about 1 mag for both sub-clusters. If we interpret this as a proxy for an age spread, then the age spreads for the sub-clusters are roughly 6 Myr, using the same assumptions as above.  An age spread in the cluster is expected: we find both regions of the cluster that have dispersed their gas and have low disk fractions as well as regions of ongoing star formation. However, there are various sources of scatter in the luminosities of YSOs that could be mistaken for an age spread. Potential sources of scatter are discussed in recent papers by \citet{jeff11} and \citet{regianni11}. If this scatter is accounted for, we would expect the age spread to be much smaller. We will examine the observed scatter in a future paper.

An examination of the disk fraction as a function of age curve from \citet{hern08} (see Fig.~\ref{fig_jesus}) shows a roughly $13 \%$ Myr$^{-1}$ drop in the disk fraction between the ages of 2 and 6 Myr.  If we assume a 1 Myr age difference between the sub-clusters, then we find a  $18 \%$ Myr$^{-1}$ drop in the disk fraction using the X-ray sample and a $10 \%$ Myr$^{-1}$ drop in the disk fraction using the visible sample.  This is roughly consistent with the disk fraction curve from \citet{hern08}.  Interestingly, if we assume that there is an age difference of 1~Myr, then the large disk fraction difference seen in the blue bin in Table  \ref{color_table} could imply that disk dissipation is more rapid in the higher mass stars.  This would be consistent with the observations of  \citet{lada06} and \citet{hern09}.

 

\section{SUMMARY\label{summary}} 

1) We have surveyed the full spatial extent of the young Cep OB3b cluster with
the IRAC and MIPS instruments onboard $\it{Spitzer}$, and revealed over 1000 young stellar objects with infrared excesses.  In addition we have performed $\it{Chandra}$ imaging of the cluster with the ACIS instrument.  Using these and archival $\it{Chandra}$ data, we find 499 X-ray bright, young stars without infrared excesses.  We find most of the cluster lies in a cavity of low extinction.   The spatial distribution of these young sources shows a hierarchical morphology composed of two large sub-clusters with a diffuse halo.  The two sub-clusters are associated with separate molecular clumps which contain ongoing star formation.

2) Using these data we constrain the location of potential Cep OB3b members in $V$ and $V - I$ color-magnitude space.  We use these constraints, in addition to the known X-ray detections, to identify the diskless Class III sources, and conduct a census of the young members of Cep OB3b.  The average disk fraction for the region is $33\% \pm 2\%$. These data, show Cep~OB3b as a massive, nearby young cluster containing about 3000 stars, depending on the number of diskless pre-MS stars. 

3) Using this census of near-IR excess and diskless sources, we create a disk fraction map of the cluster.  We find that the observed disk fraction varies spatially throughout the cluster, with a significant difference in the disk fractions of the two sub-clusters; $32\% \pm 4\%$ in the eastern sub-cluster and $50\% \pm 6\%$ in the western sub-cluster.  This difference is dominated by the sources with the bluest $V - I$ colors, corresponding to the higher mass members of the sub-clusters ($>$ 1.7~$\msol$).

4) We compare the spatial variation in the disk fraction against the location of the most massive stars in the cluster and find no clear signature of disk photoevaporation.  Further, using visible photometry, we estimate the age difference between the two sub-clusters to be as much as 1 Myr.   The different disk fractions in the sub-clusters over this age range is consistent with the rate of observed disk dissipation in other star forming regions.  We conclude that the most likely reason for the spatially varying disk fraction is a mixture of ages.

 5) We find that the
size and membership of Cep OB3b is similar to that
of the ONC, making it one of the largest, known, dense, young clusters within
1 kpc of the Sun. Most of the Cep OB3b cluster lies in a
low-extinction cavity, suggesting that Cep OB3b has largely dispersed
its natal gas and is more evolved than the more embedded ONC.  The size and evolutionary state of Cep OB3b makes it a unique target
for studying the evolution of young clusters.


\acknowledgments

This work is based on observations made with the Spitzer Space
Telescope, which is operated by the Jet Propulsion Laboratory,
California Institute of Technology, under NASA contract 1407.  Support
for this work was provided by NASA through an award issued by
JPL/Caltech.  Support for this work was provided by the National Aeronautics and Space Administration through Chandra Award Number GO9-0017A issued by the Chandra X-ray Observatory Center, which is operated by the Smithsonian Astrophysical Observatory for and on behalf of the National Aeronautics Space Administration under contract NAS8-03060.  STM and EK acknowledge support by NASA award 1281302.  JLP acknowledges support from the NASA award 1276271.  SJW and BDS were supported by NASA contract NAS8-03060 to the Chandra Science Center.  Additional support for this work was provided by the National Science Foundation award AST-1009564.  This research has made use of the NASA/IPAC Infrared Science Archive, which is
operated by the Jet Propulsion Laboratory, California Institute of
Technology, under contract with the National Aeronautics and Space
Administration. This publication makes use of data products from the
Two Micron All Sky Survey, which is a joint project of the University
of Massachusetts and the Infrared Processing and Analysis
Center/California Institute of Technology, funded by the National
Aeronautics and Space Administration and the National Science
Foundation and JPL support from SAO/JPL SV4-74011.

\begin{deluxetable}{llcccccccccccc}
\rotate
\tabletypesize{\scriptsize}
\setlength{\tabcolsep}{0.02in}
\hspace*{-1in}
\tablecolumns{14}
\tablewidth{0pc}
\tablecaption{Young Stellar Objects: Class I  \label{classItable}}
\tablehead{
\colhead{RA$_{2000}$}\tablenotemark{a} & \colhead{DEC$_{2000}$}\tablenotemark{a} & \colhead{$V$}\tablenotemark{b} & \colhead{$V - I$}\tablenotemark{b}  & \colhead{$J$}\tablenotemark{c} & \colhead{$H$}\tablenotemark{c} & \colhead{$K_s$}\tablenotemark{c} & \colhead{$[3.6]$}\tablenotemark{d} & \colhead{$[4.5]$}\tablenotemark{d} & \colhead{$[5.8]$}\tablenotemark{d} & \colhead{$[8.0]$}\tablenotemark{d} & \colhead{$[24]$}\tablenotemark{e} & \colhead{A$_{Ks}$}\tablenotemark{f} & \colhead{Notes}\tablenotemark{g}}
\startdata

22:50:14.22 &  62:25:50.2 &  \nodata &  \nodata &  \nodata &  \nodata &  15.17$\pm$0.16 &  14.07$\pm$0.01 &  13.37$\pm$0.01 &  12.66$\pm$0.02 &  11.59$\pm$0.03 &  8.82$\pm$0.08 &  0.00 &  \nodata \\ 
22:51:22.84 &  62:48:40.8 &  \nodata &  \nodata &  \nodata &  \nodata &  \nodata &  15.09$\pm$0.02 &  13.88$\pm$0.01 &  12.72$\pm$0.02 &  11.40$\pm$0.02 &  7.71$\pm$0.03 &  0.00 &  \nodata \\ 
22:52:22.88 &  62:28:46.6 &  \nodata &  \nodata &  \nodata &  \nodata &  \nodata &  14.40$\pm$0.02 &  13.33$\pm$0.01 &  12.48$\pm$0.05 &  11.30$\pm$0.10 &  6.30$\pm$0.11 &  0.00 &  \nodata \\ 
22:52:49.10 &  62:29:20.5 &  20.89$\pm$0.01 &  1.96$\pm$0.02 &  \nodata &  15.64$\pm$0.16 &  14.36$\pm$0.08 &  12.17$\pm$0.01 &  11.29$\pm$0.01 &  10.31$\pm$0.01 &  9.22$\pm$0.01 &  6.38$\pm$0.03 &  0.00 &  \nodata \\ 
22:53:02.63 &  62:33:39.4 &  24.35$\pm$0.14 &  6.68$\pm$0.16 &  15.29$\pm$0.05 &  14.20$\pm$0.07 &  13.79$\pm$0.05 &  13.77$\pm$0.05 &  13.03$\pm$0.03 &  12.20$\pm$0.04 &  10.82$\pm$0.02 &  8.39$\pm$0.05 &  0.06 &  \nodata \\ 
\enddata
\tablecomments{Only a portion of the table is shown here.  The complete table is available in the online journal.}
\tablenotetext{a}{RA and Dec values from the $\it{Spitzer}$ $[3.6]$ astrometry.}
\tablenotetext{b}{$V$ and $I$ band data from \citet{mayne07}.}
\tablenotetext{c}{$J$, $H$ and $K_{s}$ band data from 2MASS.}
\tablenotetext{d}{$\it{Spitzer}$ IRAC photometry.}
\tablenotetext{e}{$\it{Spitzer}$ MIPS photometry.}
\tablenotetext{f}{$A_{K_{s}}$ determined along the line of sight to the source using the method described in phase 2 of $\S$ \ref{identifying}.  These values can be converted to $A_{V}$ using the relation $A_V= 9.9A_{K_s}$.  If $A_{K_{s}}$ could not be estimated a value of 0.00 is shown.}
\tablenotetext{g}{Sources detected with $\it{Chandra}$ are identified in the notes column with "X-ray".}
\end{deluxetable}

\clearpage

\begin{deluxetable}{llcccccccccccc}
\rotate
\tabletypesize{\scriptsize}
\setlength{\tabcolsep}{0.02in}
\hspace*{-1in}
\tablecolumns{14}
\tablewidth{0pc}
\tablecaption{Young Stellar Objects: Class II  \label{classIItable}}
\tablehead{
\colhead{RA$_{2000}$}\tablenotemark{a} & \colhead{DEC$_{2000}$}\tablenotemark{a} & \colhead{$V$}\tablenotemark{b} & \colhead{$V - I$}\tablenotemark{b}  & \colhead{$J$}\tablenotemark{c} & \colhead{$H$}\tablenotemark{c} & \colhead{$K_s$}\tablenotemark{c} & \colhead{$[3.6]$}\tablenotemark{d} & \colhead{$[4.5]$}\tablenotemark{d} & \colhead{$[5.8]$}\tablenotemark{d} & \colhead{$[8.0]$}\tablenotemark{d} & \colhead{$[24]$}\tablenotemark{e}& \colhead{A$_{Ks}$}\tablenotemark{f} & \colhead{Notes}\tablenotemark{g}}
\startdata
22:48:33.73 &  62:48:47.9 &  \nodata &  \nodata &  15.09$\pm$0.04 &  14.46$\pm$0.06 &  14.27$\pm$0.06 &  14.04$\pm$0.01 &  13.74$\pm$0.01 &  14.15$\pm$0.13 &  \nodata &  \nodata &  0.00 &  \nodata \\ 
22:48:36.27 &  62:48:08.7 &  \nodata &  \nodata &  12.00$\pm$0.02 &  11.37$\pm$0.02 &  10.78$\pm$0.02 &  9.60$\pm$0.01 &  8.83$\pm$0.01 &  8.27$\pm$0.01 &  7.13$\pm$0.01 &  4.35$\pm$0.01 &  0.00 &  \nodata \\ 
22:48:40.12 &  62:56:56.4 &  \nodata &  \nodata &  14.17$\pm$0.02 &  13.54$\pm$0.02 &  13.24$\pm$0.02 &  13.08$\pm$0.02 &  12.69$\pm$0.02 &  12.69$\pm$0.20 &  \nodata &  \nodata &  0.00 &  \nodata \\ 
22:48:47.77 &  62:26:36.8 &  \nodata &  \nodata &  14.92$\pm$0.05 &  14.14$\pm$0.06 &  13.66$\pm$0.05 &  13.19$\pm$0.01 &  13.02$\pm$0.01 &  12.58$\pm$0.05 &  12.01$\pm$0.04 &  \nodata &  0.00 &  \nodata \\ 
22:48:48.63 &  62:26:36.6 &  \nodata &  \nodata &  14.32$\pm$0.03 &  13.23$\pm$0.04 &  12.63$\pm$0.02 &  12.13$\pm$0.01 &  11.41$\pm$0.01 &  11.03$\pm$0.01 &  10.31$\pm$0.01 &  \nodata &  0.04 &  \nodata \\

\enddata
\tablecomments{Only a portion of the table is shown here.  The complete table is available in the online journal.}
\tablenotetext{a}{RA and Dec values from the $\it{Spitzer}$ $[3.6]$ astrometry.}
\tablenotetext{b}{$V$ and $I$ band data from \citet{mayne07}.}
\tablenotetext{c}{$J$, $H$ and $K_{s}$ band data from 2MASS.}
\tablenotetext{d}{$\it{Spitzer}$ IRAC photometry.}
\tablenotetext{e}{$\it{Spitzer}$ MIPS photometry.}
\tablenotetext{f}{$A_{K_{s}}$ determined along the line of sight to the source using the method described in phase 2 of $\S$ \ref{identifying}.  These values can be converted to $A_{V}$ using the relation $A_V= 9.9A_{K_s}$.  If $A_{K_{s}}$ could not be estimated a value of 0.00 is shown.}
\tablenotetext{g}{Sources detected with $\it{Chandra}$ are identified in the notes column with "X-ray".}
\end{deluxetable}

\clearpage

\begin{deluxetable}{llcccccccccccc}
\rotate
\tabletypesize{\scriptsize}
\setlength{\tabcolsep}{0.02in}
\hspace*{-1in}
\tablecolumns{14}
\tablewidth{0pc}
\tablecaption{Young Stellar Objects: Transition Disk  \label{classtdtable}}
\tablehead{
\colhead{RA$_{2000}$}\tablenotemark{a} & \colhead{DEC$_{2000}$}\tablenotemark{a} & \colhead{$V$}\tablenotemark{b} & \colhead{$V - I$}\tablenotemark{b}  & \colhead{$J$}\tablenotemark{c} & \colhead{$H$}\tablenotemark{c} & \colhead{$K_s$}\tablenotemark{c} & \colhead{$[3.6]$}\tablenotemark{d} & \colhead{$[4.5]$}\tablenotemark{d} & \colhead{$[5.8]$}\tablenotemark{d} & \colhead{$[8.0]$}\tablenotemark{d} & \colhead{$[24]$}\tablenotemark{e} & \colhead{A$_{Ks}$}\tablenotemark{f} & \colhead{Notes}\tablenotemark{g}}
\startdata
22:48:50.00 &  62:33:25.7 &  \nodata &  \nodata &  14.85$\pm$0.03 &  13.98$\pm$0.04 &  13.60$\pm$0.03 &  13.15$\pm$0.01 &  12.95$\pm$0.01 &  12.62$\pm$0.02 &  12.42$\pm$0.07 &  8.21$\pm$0.06 &  0.02 &  \nodata \\ 
22:49:17.26 &  62:37:22.2 &  \nodata &  \nodata &  10.59$\pm$0.02 &  10.42$\pm$0.02 &  10.35$\pm$0.02 &  10.29$\pm$0.01 &  10.28$\pm$0.01 &  10.19$\pm$0.01 &  9.81$\pm$0.01 &  7.32$\pm$0.05 &  0.00 &  \nodata \\ 
22:50:44.67 &  62:45:28.2 &  \nodata &  \nodata &  14.83$\pm$0.03 &  13.80$\pm$0.04 &  13.43$\pm$0.03 &  13.13$\pm$0.01 &  13.03$\pm$0.01 &  12.95$\pm$0.02 &  12.80$\pm$0.04 &  9.58$\pm$0.18 &  0.05 &  \nodata \\ 
22:50:49.59 &  62:57:14.8 &  \nodata &  \nodata &  14.64$\pm$0.04 &  13.76$\pm$0.04 &  13.48$\pm$0.03 &  13.18$\pm$0.01 &  13.01$\pm$0.01 &  12.83$\pm$0.03 &  11.78$\pm$0.04 &  7.36$\pm$0.03 &  0.03 &  \nodata \\ 
22:50:49.95 &  62:31:41.1 &  \nodata &  \nodata &  12.39$\pm$0.02 &  11.96$\pm$0.02 &  11.80$\pm$0.02 &  11.73$\pm$0.01 &  11.68$\pm$0.01 &  11.67$\pm$0.01 &  11.61$\pm$0.04 &  8.37$\pm$0.07 &  0.00 &  \nodata \\ 

\enddata
\tablecomments{Only a portion of the table is shown here.  The complete table is available in the online journal.}
\tablenotetext{a}{RA and Dec values from the $\it{Spitzer}$ $[3.6]$ astrometry.}
\tablenotetext{b}{$V$ and $I$ band data from \citet{mayne07}.}
\tablenotetext{c}{$J$, $H$ and $K_{s}$ band data from 2MASS.}
\tablenotetext{d}{$\it{Spitzer}$ IRAC photometry.}
\tablenotetext{e}{$\it{Spitzer}$ MIPS photometry.}
\tablenotetext{f}{$A_{K_{s}}$ determined along the line of sight to the source using the method described in phase 2 of $\S$ \ref{identifying}.  These values can be converted to $A_{V}$ using the relation $A_V= 9.9A_{K_s}$.  If $A_{K_{s}}$ could not be estimated a value of 0.00 is shown.}
\tablenotetext{g}{Sources detected with $\it{Chandra}$ are identified in the notes column with "X-ray".}
\end{deluxetable}

\clearpage

\begin{deluxetable}{llcccccccccccc}
\rotate
\tabletypesize{\scriptsize}
\setlength{\tabcolsep}{0.02in}
\hspace*{-1in}
\tablecolumns{14}
\tablewidth{0pc}
\tablecaption{Young Stellar Objects: Class III  \label{classtdtable}}
\tablehead{
\colhead{RA$_{2000}$}\tablenotemark{a} & \colhead{DEC$_{2000}$}\tablenotemark{a} & \colhead{$V$}\tablenotemark{b} & \colhead{$V - I$}\tablenotemark{b}  & \colhead{$J$}\tablenotemark{c} & \colhead{$H$}\tablenotemark{c} & \colhead{$K_s$}\tablenotemark{c} & \colhead{$[3.6]$}\tablenotemark{d} & \colhead{$[4.5]$}\tablenotemark{d} & \colhead{$[5.8]$}\tablenotemark{d} & \colhead{$[8.0]$}\tablenotemark{d} & \colhead{$[24]$}\tablenotemark{e} & \colhead{A$_{Ks}$}\tablenotemark{f} & \colhead{Notes}\tablenotemark{g}}
\startdata
22:51:55.39 &  62:41:44.7 &  \nodata &  \nodata &  11.64$\pm$0.02 &  10.79$\pm$0.02 &  10.51$\pm$0.01 &  10.34$\pm$0.01 &  10.32$\pm$0.01 &  10.23$\pm$0.01 &  10.21$\pm$0.01 &  9.97$\pm$0.10 &  0.03 &  X-ray \\ 
22:52:16.16 &  62:33:01.7 &  \nodata &  \nodata &  12.84$\pm$0.02 &  12.10$\pm$0.02 &  11.80$\pm$0.02 &  11.60$\pm$0.01 &  11.55$\pm$0.01 &  11.51$\pm$0.01 &  11.50$\pm$0.03 &  \nodata &  0.01 &  X-ray \\ 
22:52:21.11 &  62:44:47.2 &  18.28$\pm$0.01 &  2.64$\pm$0.01 &  13.74$\pm$0.02 &  12.79$\pm$0.02 &  12.51$\pm$0.02 &  12.30$\pm$0.01 &  12.33$\pm$0.01 &  12.22$\pm$0.01 &  12.17$\pm$0.03 &  \nodata &  0.04 &  \nodata \\ 
22:52:22.12 &  62:34:54.3 &  19.99$\pm$0.01 &  3.26$\pm$0.01 &  14.31$\pm$0.04 &  13.22$\pm$0.04 &  \nodata &  12.53$\pm$0.01 &  12.50$\pm$0.01 &  12.37$\pm$0.02 &  12.38$\pm$0.05 &  \nodata &  0.00 &  \nodata \\ 
22:52:22.27 &  62:30:33.0 &  18.88$\pm$0.01 &  2.81$\pm$0.01 &  13.96$\pm$0.02 &  12.96$\pm$0.02 &  12.62$\pm$0.02 &  12.34$\pm$0.01 &  12.37$\pm$0.01 &  12.28$\pm$0.01 &  12.31$\pm$0.05 &  \nodata &  0.05 &  \nodata \\ 

\enddata
\tablecomments{Only a portion of the table is shown here.  The complete table is available in the online journal.}
\tablenotetext{a}{RA and Dec values from the $\it{Spitzer}$ $[3.6]$ astrometry.}
\tablenotetext{b}{$V$ and $I$ band data from \citet{mayne07}.}
\tablenotetext{c}{$J$, $H$ and $K_{s}$ band data from 2MASS.}
\tablenotetext{d}{$\it{Spitzer}$ IRAC photometry.}
\tablenotetext{e}{$\it{Spitzer}$ MIPS photometry.}
\tablenotetext{f}{$A_{K_{s}}$ determined along the line of sight to the source using the method described in phase 2 of $\S$ \ref{identifying}.  These values can be converted to $A_{V}$ using the relation $A_V= 9.9A_{K_s}$.  If $A_{K_{s}}$ could not be estimated a value of 0.00 is shown. }
\tablenotetext{g}{Sources detected with $\it{Chandra}$ are identified in the notes column with "X-ray".}
\end{deluxetable}

\clearpage

\begin{deluxetable}{lccccc}
\tabletypesize{\scriptsize}
\setlength{\tabcolsep}{0.02in}
\hspace*{-1in}
\tablecolumns{14}
\tablewidth{0pc}
\tablecaption{Disk Fractions in Cep OB3b  \label{df_table}}
\tablehead{
\colhead{Method}\tablenotemark{a} & \colhead{Total}\tablenotemark{b} & \colhead{Overlap}\tablenotemark{c} & \colhead{Eastern}\tablenotemark{d}  & \colhead{Western}\tablenotemark{e}  & \colhead{Notes}\tablenotemark{f}
}
\startdata
Visible & $25\%\pm1\%$ & $30\%\pm2\%$ & $35\%\pm3\%$ & $45\%\pm5\%$  & \nodata \\
 & $425/(425+1270)$\tablenotemark{g} & $389/(389+914)$\tablenotemark{g} & $121/(121+222)$ \tablenotemark{g} & $87/(87+106)$\tablenotemark{g}   &  \\
X-ray  & $33\%\pm2\%$  & $33\%\pm2\%$  & $32\%\pm4\%$ & $50\%\pm6\%$  &  \nodata  \\
 & $254/(254+499)$\tablenotemark{g} & $241/(241+472)$\tablenotemark{g}   & $74/(74+151)$ \tablenotemark{g} & $80/(80+79)$\tablenotemark{g}  &  \\
\hline
Visible & $32\%\pm2\%$ & $36\%\pm2\%$ & $39\%\pm4\%$ & $53\%\pm6\%$  & Sub \\
 & $425/(425+889)$\tablenotemark{g} & $389/(389+914)$\tablenotemark{g} & $121/(121+191)$ \tablenotemark{g} & $87/(87+75)$\tablenotemark{g}   &  \\
X-ray  & $36\%\pm2\%$  & $36\%\pm2\%$   & $34\%\pm4\%$ & $52\%\pm6\%$ &  Sub  \\
 & $254/(254+446)$\tablenotemark{g} & $241/(241+431)$\tablenotemark{g}  & $74/(74+145)$ \tablenotemark{g} & $80/(80+73)$\tablenotemark{g}   &  \\
\hline
Visible & $24\%\pm1\%$ & $29\%\pm2\%$ & $34\%\pm3\%$ & $44\%\pm5\%$  & Cut \\
 & $367/(367+1139)$\tablenotemark{g} & $338/(338+811)$\tablenotemark{g}  & $103/(103+199)$\tablenotemark{g} & $75/(75+96)$\tablenotemark{g} &  \\
X-ray  & $33\%\pm2\%$  & $33\%\pm2\%$   & $32\%\pm4\%$\tablenotemark{h}  & $50\%\pm6\%$\tablenotemark{h} & Cut \\
 & $223/(223+451)$\tablenotemark{g} & $214/(214+426)$\tablenotemark{g}  &  $64/(64+137)$\tablenotemark{g} & $71/(71+71)$ \tablenotemark{g} &  \\
\enddata
\tablenotetext{a}{Method used to determine potential Class III members.}
\tablenotetext{b}{The full field of view for the given sample.}
\tablenotetext{c}{The field of view common to both the visible and the X-ray sample.}
\tablenotetext{d}{$5'$ radius regions centered on RA: 22:56:49.44, Dec: 62:39:55.54.}
\tablenotetext{e}{$5'$ radius regions centered on RA: 22:53:50.83, Dec: 62:35:36.37.}
\tablenotetext{f}{Samples with estimated contaminants subtracted are identified with the note Sub.  See $\S$ \ref{xr_ind} for details about the X-ray contamination and $\S$ \ref{df_vis} for details about the visible contamination.  Samples with the NIR color and magnitude cut applied are identified with the note Cut.  See $\S$~\ref{sub_fractions} for details of the NIR color and magnitude cuts.}
\tablenotetext{g}{These numbers are (number of disk sources)/(number of disk sources + number of non-disk sources)}
\tablenotetext{h}{These are the disk fractions quoted in the abstract.}
\end{deluxetable}

\clearpage

\begin{deluxetable}{lcccccc}
\tabletypesize{\scriptsize}
\setlength{\tabcolsep}{0.02in}
\hspace*{-1in}
\tablecolumns{14}
\tablewidth{0pc}
\tablecaption{Disk Fractions in the Sub-Clusters as a Function of Dereddened $V - I$ color  \label{color_table}}
\tablehead{
\colhead{Color Range}\tablenotemark{a} & \colhead{Spectral Type}\tablenotemark{b}  & \colhead{Mass}\tablenotemark{c} & \colhead{Sample}\tablenotemark{d} & \colhead{Overlap}\tablenotemark{e}  & \colhead{Eastern}\tablenotemark{f} & \colhead{Western}\tablenotemark{g}
}
\startdata
-0.5 to 1.05 mag & $>$ K3 & $>$ 1.7 & Visible & $19\%\pm2\%$ &  $19\%\pm4\%$ & $39\%\pm9\%$  \\
& & & & $72/(72+311)$\tablenotemark{h} &  $18/(18+78)$\tablenotemark{h}  & $20/(20+31)$\tablenotemark{h}  \\

 & && Visible Subtracted & $30\%\pm4\%$ &  $23\%\pm6\%$ & $63\%\pm14\%$  \\
& & & & $72/(72+172)$\tablenotemark{h} &  $18/(18+59)$\tablenotemark{h}  & $20/(20+12)$\tablenotemark{h}  \\
\hline
& & & X-ray  &  $30\%\pm5\%$  &  $15\%\pm6\%$ & $56\%\pm14\%$   \\
& & & &  $44/(44+103)$ \tablenotemark{h}  & $8/(8+43)$\tablenotemark{h}  & $15/(15+12)$\tablenotemark{h}  \\
\hline
\hline
1.05 to 1.72 mag & K3 to M0 & 1.7 to 0.6 & Visible  & $31\%\pm3\%$  &  $40\%\pm6\%$ & $39\%\pm9\%$  \\
& & & &  $120/(120+263)$\tablenotemark{h} & $40/(40+59)$\tablenotemark{h} & $25/(25+39)$\tablenotemark{h}   \\

 &  &  & Visible Subtracted  & $38\%\pm3\%$  &  $44\%\pm7\%$ & $45\%\pm9\%$  \\
& & & &  $120/(120+199)$\tablenotemark{h} & $40/(40+50)$\tablenotemark{h} & $25/(25+30)$\tablenotemark{h}   \\
\hline
& & & X-ray &  $32\%\pm4\%$  & $33\%\pm7\%$ & $46\%\pm11\%$    \\
&& & & $58/(58+124)$\tablenotemark{h} &  $21/(21+42)$\tablenotemark{h} & $17/(17+22)$\tablenotemark{h} \\
 \hline
 \hline
1.72 to 4.35 mag & M0 to M6 & 0.6 to 0.2 & Visible  & $38\%\pm3\%$ & $42\%\pm6\%$  & $54\%\pm10\%$  \\
&  & & & $146/(146+237)$\tablenotemark{h}  & $45/(45+62)$\tablenotemark{h} & $30/(30+26)$\tablenotemark{h}   \\ 

 & &  & Visible Subtracted  & $41\%\pm3\%$ & $43\%\pm7\%$  & $57\%\pm10\%$  \\
&  & & & $146/(146+237)$\tablenotemark{h}  & $45/(45+59)$\tablenotemark{h} & $30/(30+23)$\tablenotemark{h}   \\ 
\hline
& & & X-ray   &  $31\%\pm5\%$  &  $26\%\pm8\%$ & $50\%\pm16\%$   \\
 & & & &  $36/(36+80)$\tablenotemark{h} & $10/(10+28)$\tablenotemark{h}  & $10/(10+10)$\tablenotemark{h} \\
\enddata
\tablecomments{Due to the unknown $V - I$ color dependence of the estimated X-ray contamination, only the visible sample has contamination subtracted.}
\tablenotetext{a}{The total sample was divided into blue, intermediate and red thirds in the dereddened $V - I$ color space.  The colors given are the bounding colors.  Also given is the corresponding spectral type and corresponding mass range for these bounding colors.  See $\S$ \ref{color_df} for details.}
\tablenotetext{b}{Spectral types correspond to the bounding colors are from \citet{kenyon1995}.}
\tablenotetext{c}{Masses corresponding to the bounding colors are estimated fromthe models of \citet{siess00}.}
\tablenotetext{d}{Sample used to determine potential Class III members.}
\tablenotetext{c}{All sources in the field of view common to both samples.}
\tablenotetext{d}{$5'$ radius region centered on: RA: 22:56:49.44, Dec: 62:39:55.54.}
\tablenotetext{e}{$5'$ radius region centered on: RA: 22:53:50.83, Dec: 62:35:36.37.}
\tablenotetext{h}{These numbers are (number of disk sources)/(number of disk sources + number of non-disk sources)}
\end{deluxetable}

\clearpage

\end{document}